\documentclass[useAMS,usenatbib]{mn2e}

\usepackage{amsmath}
\usepackage{url}
\usepackage{amsfonts}
\usepackage{amsbsy}
\usepackage{graphicx}
\usepackage{subfigure}
\usepackage{verbatim}
\usepackage{amssymb}

\renewcommand{\u}{\ensuremath{\mathbf{u}}}
\renewcommand{\d}{\ensuremath{\partial}}

\newcommand{\ey}{\ensuremath{\mathbf{e}_{y}}}
\newcommand{\ez}{\ensuremath{\mathbf{e}_{z}}}
\newcommand{\ephi}{\ensuremath{\mathbf{e}_{\phi}}}
\newcommand{\wt}{\ensuremath{\widetilde{\omega}}}
\newcommand{\beq}{\begin{equation}}                        
\newcommand{\eeq}{\end{equation}}

\begin{document}

\title[Inertial waves in 3D disks]{Inertial waves near corotation in 3D
hydrodynamical disks}
\author[Henrik N. Latter and Steven A. Balbus]
{Henrik N. Latter\thanks{E-mail: henrik.latter@lra.ens.fr}
 and Steven A. Balbus\thanks{E-mail: steven.balbus@lra.ens.fr}\\
Laboratoire de Radioastronomie,
 \'Ecole Normale Sup\'erieure,
 24 rue Lhomond, Paris 75005,  France}

\maketitle

\begin{abstract}
This paper concerns the interaction
 between non-axisymmetric inertial
waves and their corotation resonances in a
hydrodynamical disk.
Inertial waves are of interest because they 
can localise in resonant cavities circumscribed by Lindblad radii, and as
 a consequence exhibit
discrete oscillation frequencies that may be observed. It is often
 hypothesised that these trapped eigenmodes are
 affiliated with the poorly understood QPO phenomenon.
 We demonstrate that a large class of non-axisymmetric 3D inertial waves
 cannot manifest as trapped normal modes. This class includes any
 inertial wave whose
 resonant cavity contains a corotation singularity. Instead, these
 `singular' modes
 constitute a continuous spectrum and, as an ensemble, are convected with
 the flow, giving rise to shearing
 waves. Lastly, we present a simple demonstration of how the corotation
 singularity stabilizes three-dimensional perturbations in a slender torus.
\end{abstract}

\section{Introduction}

Inertial waves, also called r-modes (Korycanksy and Pringle 1995, Ogilvie
 1998)  or g-modes (Wagoner 1999, Kato
2001a), are one of a rich
 assortment
 of oscillations exhibited by hydrodynamical models of
accretion disks. They may be distinguished by a number of
interesting
 properties:
 first, they comprise
motions that are effectively incompressible, and which
strongly couple the vertical and horizontal
velocities.
Second, their oscillation frequencies take values much less than the local
epicyclic frequency, which suggests that inertial waves figure prominently
in a disk's response to low frequency forcing (Balbus 2003), such as
might issue from an embedded protoplanet.
Third, these modes are the only waves that propagate in the
vicinity of their
corotation resonances, i.e.\ the radii where their pattern speed is
zero. Finally, in principle they may become trapped between their
two Lindblad resonances and hence manifest as a discrete suite of standing modes
(see, for
example, Okazaki et al.~1987, Perez et al.~1997).

Inertial waves have been invoked to explain various astrophysical
phenonomena, but of special interest is their potential role in the
excitation of the quasi-periodic oscillations (QPOs) that are observed in
the X-ray spectra of some black-hole disk systems (e.g., Wagoner 1999;
Kato 2001a, McClintock and Remillard 2003).  The idea is that since
the inertial modes appear as trapped standing waves, they (or their
resonant interactions) may exhibit oscillation frequencies in accord
with the observed frequencies. Meanwhile, a variety of instabilities have
 been promoted in order to explain the large amplitudes necessary for
the detection of these frequencies (Abramowicz and Klu\'{z}niak 2001, Kato 2001b, 2003b,
Ferreira and Ogilvie 2008, for example).

Though axisymmetric waves (Okazaki et al.~1987) and certain
non-axisymmetric waves (Perez et al.~1997) have been demonstrated
mathematically to form discrete trapped standing waves, this behaviour has
not been established satisfactorily for general three-dimensional modes.
In this paper we show, with both WKBJ techniques and direct methods, that
in most cases, three-dimensional inertial waves do not manifest this way. This
is because  
most non-axisymmetric 3D modes possess a corotation singularity within
their resonant cavity.  While earlier work (Kato 2003a, Li et al.~2003)
has shown how the corotation singularity damps incident wave fluxes, it
has not been generally appreciated that the corotation singularity
also very likely \emph{prohibits} the formation of trapped normal modes
themselves.  The present paper is concerned with investigating
the dynamics of this problem.  In addition, we study
the continuous spectrum that issues from the corotation singularity,
showing its close relationship with inertial shearing waves (Johnson and
Gammie 2005, Balbus and Hawley 2006). Lastly, we demonstrate how
the wave absorption at corotation impedes unstable three-dimensional
modes in a slender torus. In so doing, we deepen our understanding of
the complicated and varied linear responses available to accretion disks.

The outline of this paper is as follows.  Section 2 presents the
governing equations of an incompressible disk in a shearing sheet and
summarises the main qualitative features of travelling inertial waves
near the corotation point.  Section 3 demonstrates the absence of trapped
modes around corotation by solving the boundary value problem. In the
following section we provide details of a WKBJ analysis which supplies
the same result in a more realistic semi-global model but in the limit of
large vertical wavenumber.  Section 5 outlines the relationship between
shearing waves and the continuous spectrum induced by the corotation
singularity, while Section 6 describes instability in a 3D slender torus.
Our conclusions are drawn in Section 7.

\section{Preliminaries}

Our treatment is restricted to inertial waves whose resonant cavities
contain the corotation singularity. Such a configuration is typical for 3D
inertial waves in both Newtonian and (general) relativistic disks, 
with only
a narrow band of non-axisymmetric r-modes in relativistic disks exhibiting
a different configuration of resonances (see Perez et al.~1997,
Wagoner 1999, Kato 2001a, 2001b).  The more common modes were first
studied in detail by Kato (2002), who suggested that the corotation
point could give rise to a form of the Papaloizou-Pringle instability.
Subsequent WKBJ and numerical analyses, however, demonstrated that an
incident inertial wave would be strongly absorbed at corotation
(Kato 2003a, Li et al. 2003, but see Drury, 1985, for a
much earlier discussion).  These results convinced researchers that
such trapped 3D modes must decay.  As a consequence, the emphasis of
much `diskoseismology' has moved on to other mechanisms of instability,
typically involving a parametric resonance (Abramowicz and Klu\'{z}niak
2001, Kato 2003b, Ferreira and Ogilvie 2008).  However,
a central question remains unaddressed: 
 How is it possible that trapped standing
waves (growing or decaying) form at all in the vicinity of a strong wave absorber,
such as the corotation singularity?

The WKBJ analysis of Kato (2002, 2003a; hereafter K02 and K03a), assumes
the existence of such trapped modes and computes their damping rates. The
theory incorporates neither explicit nor implicit dissipation, which
allows a growing mode for each decaying mode present,
due to time symmetry.  But because the problem exhibits a branch cut,
this symmetry must be broken in any calculation.
 At the outset, one must assume either
growth or decay of the modes to be studied.  Henceforth, care must
be taken to assure self-consistency.   Both K02 and K03a assumed that
their modes grow in their calculation.  Their final result, however,
was that the modes decay.  This was treated, not as an inconsistency to
be rectified, but as indication the modes do, in fact, decay.  It is the
view of the authors that the correct interpretation of the inconsistency
is that there exists no trapped modes at all. The only class of modes
that could in principle resolve the contradiction are strictly neutral.
But such modes are singular, possessing undefined radial derivatives at
the critical point.

In Section 4 we revisit the analysis of K02 and K03a and reestablish the
above result in some detail. Before doing so, we demonstrate the main ideas
using a simple local model, the shearing sheet, and very straighforward
techniques. In the shearing sheet limit, the trapped wave problem can be
solved analytically, and hence is more readily understood.  This limit
neglects global phenomena such as curvature terms and edge effects, as
well as general relativity, but it includes all the essential physics:
a corotation singularity with two Lindblad resonances on either side.

We begin with the basic equations in the shearing sheet limit, and then
give a brief summary of travelling inertial waves. This will help us
interpret the results of the following section in which the absence of
normal modes is demonstrated.

\subsection{Governing equations}

The shearing sheet (Goldreich and Lynden-Bell 1965) is designed to
study local disk behavior.  We make the additional assumptions that
self-gravity is unimportant, and that the fluid is incompressible.
Also, we work in the `cylindrical limit': the equilibrium density is
unstratified, independent of both cylindrical radius $(r)$ and height
($z$).  It can be shown, however, that the vertical stratification plays no role
in the qualitative behaviour of our problem. For example, an isothermal
model yields analogous results, at greater mathematical cost (see
Appendix A).

The governing
equations in the shearing sheet read:
\begin{align} \label{st1}
\d_t \u + \u \cdot\nabla \u + 2\Omega_0\ez \times \u & = -\nabla \Phi
-\frac{1}{\rho}\nabla p, \\
  \nabla\cdot \u &=0 ,\label{st2}
\end{align}
where $\u$ is velocity, $\Phi$ is the tidal potential, $p$ is pressure, and
$\rho$ is density (assumed a constant). Additionally,
$\Omega_0\equiv\Omega(r_0)$ is the local orbital frequency of the disk, where
$r_0$ is the radius at which the shearing sheet is anchored. The $x$-coordinate
represents radius ($dx=dr$), the $y$ coordinate the azimuthal
direction ($dy= rd\theta$), and the $z$ coordinate the vertical direction. The
sheet undergoes a background linear shear flow
\begin{equation}
\u_0= -q\Omega_0\,x\,\ey
\end{equation}
where $q= - (d\ln\Omega/d \ln r)_0.$ If the disk is Keplerian, $q=3/2$.
Henceforth, the $0$ subscript on $\Omega_0$ will be suppressed.

\subsection{Linear perturbation equations}

The governing equations admit by construction the homogeneous equilibrium $\u=\u_0$,
$\rho=\rho_0$, and $p=p_0$. Let us now introduce a small disturbance
so that $ \u=\u_0+\u' $, and $p=p_0+p'$.
The equations 
governing the linear evolution of such a pertubation are:
\begin{align} \label{lin1}
\d_t u_x' -q\Omega x\,\d_y u_x' -2\Omega u_y' &= -\d_x h', \\
\d_t u_y' -q\Omega x\, \d_y u_y' +\tfrac{1}{2}(\kappa^2/\Omega)u_x' &= -\d_y h', \\
\d_t u_z'-q\Omega x\,\d_y u_z' &= -\d_z h', \label{lin3}
\end{align}
with the incompressibility restriction:
\begin{equation}\label{lin4}
\d_x u_x'+\d_y u_y' + \d_z u_z' = 0.
\end{equation}
Note that the pertubed enthalpy $h'=p'/\rho_0$ takes the place of the perturbed
pressure $p'$.
 The epicyclic frequency is defined through
$ \kappa^2=  2(2-q)\Omega^2.$

The perturbations are Fourier decomposed in space so that each mode
is $\propto \text{exp}( i k_y y+ i k_z z)$. The
adoption of sinusoidal variation in $z$ is permitted by the cylindrical
approximation. After some manipulation we arrive upon the central equation of the
linear
analysis for a single mode $(k_y,k_z)$,
\begin{equation} \label{firststep}
(\d_t-iq\Omega x k_y)^2(\d_x^2-k_\perp^2)u_x'+\kappa^2k_z^2 u_x'=0,
\end{equation}
where $k_\perp^2\equiv k_y^2+k_z^2$, and $u_x'$ now
represents a Fourier amplitude. At this point we perform a temporal Fourier transform
and assume the equation admits discrete modal solutions, i.e. we take
 $u_x'\propto e^{-i\omega t}$, where $\omega$ is a (complex)
 frequency. We set
$$ \omega= \omega_r + i\sigma,$$
for $\omega_r$ and $\sigma$ real. The sign of $\sigma$
must be specified now so as to negotiate the
branch point that arises from the inviscid dynamics. Without loss of
generality we set $\sigma \geq 0$, as the inviscid problem supports either neutral
modes or growing/decaying pairs (due to the time symmetry).
Equation \eqref{firststep} is now
\begin{equation} \label{Bessel1}
\widetilde{\omega}^2\,\d_x^2 u_x' + (\kappa^2 k_z^2 - k_{\perp}^2
\widetilde{\omega}^2) u_x' = 0,
\end{equation}
where
\begin{equation}
\widetilde{\omega} = \omega+  q\Omega\, x\, k_y.
\end{equation}

\subsection{Wave-shape equation}

In order to achieve a simpler version of our working equation,
 a shifted dimensionless space variable is adopted,
\begin{equation}\label{coord}
 x^*= k_\perp \left( x + \frac{\omega_r}{q\,\Omega\,k_y} \right).
\end{equation}
Hence we anchor the shearing sheet upon the radius in the disk at which the
mode is convected by the shear flow.
The star is dropped hereafter for notational convenience, in addition to the
subscript $x$ and
prime on $u_x'$. The central wave-shape equation of the analysis is consequently
\begin{equation} \label{Bessel2}
\d_x^2 u + \left[\frac{\beta^2}{(x-ix_c)^2}\,-\,1 \right]u = 0,
\end{equation}
where 
\begin{equation} x_c = -\frac{\sigma}{q\Omega k_y}. \end{equation}
It will be assumed, without loss of generality, that $k_y>0$ and so $x_c<0$.
The dimensionless parameter $\beta$ is defined through
\begin{equation}\label{beta}
 \beta=\left|\frac{\kappa\,(k_z/k_y)}{q\Omega}\right|.
\end{equation}
In fact, $\beta^2$ is directly analogous to the Richardson number (Booker and
Bretherton, 1967, and see Li et al., 2003), and measures the ratio of the
stabilizing effects of rotation, embodied in the term $\kappa^2
(k_z^2/k_y^2)$, against the destabilising effects of shear.

The governing wave-shape equation possesses a singularity at
$x= ix_c$ (the corotation point).  However, the solutions are regular
for real $x$ provided that $x_c\neq 0$, meaning that the solutions must grow (or decay).
 Strictly neutral modes (for which $x_c=0$) are undefined because
they have singular derivatives.  With $\sigma>0$, and $x_c<0$, the
singularity always lies beneath the real axis, which serves as our
contour of integration. 
Thus the branch cut extending to the
left of the singularity always lies beneath the integration contour.

The singularity corresponds to the
corotation resonance, and arises because the fluid is inviscid; it vanishes
when dissipation is added.  Note also that the singularity
depends on $k_z \neq 0$ (from Eq.\eqref{beta}), and is not present for
two-dimensional modes.  The two Lindblad resonances occur at
$$ x= \pm \beta + i x_c,$$ which are arranged symmetrically around the
corotation radius.  (This symmetry is broken when curvature effects
are included.)  The resonant cavity is defined by
$|x|\leq\beta$.  Density waves can propagate outside of this region,
but because the fluid is incompressible, the `evanescent zones' extend
to positive and negative infinity.

Trapped standing waves must decay far from
corotation in the evanescent zones: $|u|\to 0$ when $|x|\to\infty$.
 Equation \eqref{Bessel2} is then a one-dimensional eigenvalue problem
with eigenvalue $\sigma$. In fact, Eq.~\eqref{Bessel2} is a form of the Bessel
equation
and so the general solution can be written in closed form immediately
\begin{equation}\label{bes}
 u= a_1\,\sqrt{x-ix_c}\, I_{\nu}(x-ix_c) + b_1\,\sqrt{x-ix_c}\,I_{-\nu}(x-ix_c) 
\end{equation}
where $a_1$ and $b_1$ are constants,  $I_\nu(z)$ is a modified Bessel function of
first kind,
and the order $\nu$ is defined through
\begin{equation}\label{nu}
 \nu=\frac{1}{2}\sqrt{1-4\beta^2},
\end{equation}
(see, for example, Abramowitz and Stegun, 1972).
This is particularly advantageous as, unlike other work,
 we need not resort to WKBJ or Frobenius
approximations.
 The former corresponds to the limit of large imaginary order $\nu$ (i.e.\
 $\beta\gg 1$)
and the latter to small argument $x$. Two classes of solution can be
distinguished. When $\beta>1/2$ (corresponding to modes with greater
wavevector pitch $k_z/k_y$) the order of the Bessel functions $\nu$ is
imaginary and strongly oscillatory solutions are obtained. When $\beta \leq
1/2$ the profiles resemble hyperbolic trigonometric functions.

Before we attack the boundary value problem
we discuss free travelling waves within the
resonant cavity. This will illuminate some of the properties of the
standing waves near corotation and help us understand in more physical terms 
the results of Section 3.
Some of this analysis also appears in Booker and Bretherton (1967), Vishniac and
Diamond (1989), and Li et al. (2003).

\subsection{Free inertial waves near corotation}

We first discuss the properties of wave-like solutions to \eqref{Bessel1}
far from corotation $\widetilde{\omega}=0$ by making the familiar WKBJ
ansatz, $ u= R(x) e^{i \psi(x)}, $ where $\psi$ is a phase function that
varies rapidly with $x$ in comparison with $R$. We make the identification
$k_x\equiv \d_x \psi$ and also assume that $k_y\sim k_z\sim k_x$.
 This ansatz is substituted into \eqref{Bessel1} and we collect
terms of order $k_x^2$. These assumptions lead to the dispersion relation for
inertial waves,
\begin{equation}
\widetilde{\omega}^2 = \frac{k_z^2}{k^2}\,\kappa^2,
\end{equation}
where $k^2=k_x^2+k_y^2+k_z^2$ (see also Balbus 2003). 
The group velocity of these waves with respect to the
Doppler-shifted frequency $\widetilde{\omega}$ is given by
\begin{equation} \label{Groupv}
\mathbf{V}= \nabla_{\mathbf{k}}\widetilde{\omega}= 
\frac{\widetilde{\omega}}{k^2}\,\left(-k_x,\,-k_y,\, (k_x^2+k_y^2)/k_z \right),
\end{equation}
where $\mathbf{k}=(k_x,k_y,k_z)$.
Clearly, $\mathbf{k}\cdot\mathbf{V}=0$ and the group velocity is
perpendicular to the phase velocity. Moreover, the sign of the radial component of
the group velocity $V_x$
depends on both
$\mathbf{k}$ and the sign of $\widetilde{\omega}$, and consequently the
location of the wave relative to the corotation point.
 Inside corotation ($\widetilde{\omega}<0$),
 the group velocity points in the \emph{same} radial direction as that of the phase
velocity.
 But, outside corotation ($\widetilde{\omega}>0$), the group
velocity points in the \emph{opposite} radial direction to the phase velocity.

Next we treat the fate of the inertial waves as they approach corotation,
  assuming that the group velocity calculated above does not deviate
  significantly in this region. Near
  this radius a Frobenius solution may be employed. From
  Eq.~\eqref{bes}, to leading order the approximation is
\begin{equation} \label{Frobbing}
 u = a_1\,(x-ix_c)^{1/2+\nu}+ b_1\,(x-ix_c)^{1/2-\nu},
\end{equation}
where $a_1$, $b_1$ and $\nu$ are the same quantities introduced earlier.
 We adopt the WKBJ approximation and thus assume $\beta$ is large; now
 $\nu\approx i\beta$.
 Consequently, a local radial wavenumber can be defined for each of the two
solution components above.   In particular,
\beq
u = \sqrt{x-ix_c}\left(
a_1e^{i\beta\ln|x-x_c|}
+   b_1e^{-i\beta\ln|x-x_c|}
\right).
\eeq
The $(x-x_c)^{i\beta}$ component possesses $k_x=\beta/x$, and consequently 
we identify the $a_1$ function as a left-going wave
(moving from positive to negative $x$).  This is because when $x>0$ then
$k_x>0$, and so, according to Eq.~\eqref{Groupv}, the radial component of the group
velocity $V_x$ is negative.   
When $x<0$, then $k_x<0$, and $V_x$ will also be negative. A similar argument
establishes that the $b_1$ function is the wave propagating to the
right (moving from negative to positive $x$).

When one of these wave components passes through corotation, it will 
be strongly absorbed, suffering an amplitude damping proportional to
$e^{-\pi\beta}$. This is clear from the analytic continuation of the Frobenius
solution Eq.~\eqref{Frobbing} through the corotation region, noting that the
singularity $x=ix_c$ lies beneath the real line, and thus beneath the
integration contour. For large $\beta$, i.e.\ large wavevector pitch
$k_z/k_y$,
the incident waves barely penetrate
 the corotation barrier.
For general $\beta$, when each wave passes through
corotation from its proper direction, its amplitude will acquire a factor
\begin{equation}\label{phaser}
\text{exp}(\tfrac{1}{2}\,i\pi+ i\pi\nu),
\end{equation} 
from \eqref{Frobbing}. For $\beta\le 1/2$, the alteration corresponds
to a phase shift of $\pi( \nu+1/2)$.  But it nevertheless gives rise to a damping of
angular momentum and energy flux through the corotation point
(Li et al.~2003).

It is this damping that has led researchers to discount the possibility
of instability in this context: the corotation resonance is a powerful
barrier against the communication of wave motions, plundering energy from
potentially growing modes.  We now show that this barrier not only
precludes instability, it precludes standing modes entirely.

\section{Absence of trapped 3D inertial waves in the shearing sheet model}

In this section we take the exact solution \eqref{bes} to the governing
wave-mode equation \eqref{Bessel2} and show that it cannot satisfy the
two decaying boundary conditions at $|x|\to\infty$. Physically, the reason for
this is that the standing mode consists of a precise balance of two
traveling waves (the left and right-going WKBJ waves of the previous
section) and the corotation point will heavily damp one component relative to the
other.  Consequently, the correct balance of components on one side of
corotation will never correspond to the correct balance on the other side of corotation.

Mathematically, the result is straightforward. The boundary condition at
$x\to\infty$ forces $u$ to be proportional to the Bessel function of second kind
$K_\nu(x-ix_c)$ for all $x$, as long as $x_c\neq 0$. 
This function, however, diverges when $x\to-\infty$ and so no solution exists.

Even though the mathematics is clear, 
we develop this problem in some detail because it sheds light on
the physics of the mode prohibition. In particular, it shows that
 the problem issues from the fundamental interaction
between a wave and its corotation singularity, and is thus quite general ---
 not an artefact of the shearing
sheet model. 
After all, the shearing sheet also prohibits $k_z=0$ modes (for
which there is no
singularity) because of its simple linear shear (by analogy
with Case 1960).
Moreover, the lengthier treatment below helps us deal with the nontrivial
$x_c=0$
case, which sets up a treatment of `singular modes'
 and the continuous spectrum.

\subsection{Boundary conditions}

We apply the boundary condition $|u|\to 0$ as $|x|\to \infty$ upon
the general solution to \eqref{Bessel2}, \beq u= a_1\,\sqrt{x-ix_c}\,
I_{\nu}(x-ix_c) + b_1\,\sqrt{x-ix_c}\,I_{-\nu}(x-ix_c).\eeq Normally these
constraints would lead to two algebraic equations for $b_1/a_1$ and $x_c$.
Determining the latter is the `eigenvalue problem', or `quantisation
condition'.  As we shall see, such an equation does not appear in this problem.

Throughout this section, we take $\beta$ large (WKBJ waves) and so make
the approximation $\nu\approx i\beta$.  Each solution component above
can then be regarded as a travelling wave: the $I_{i\beta}$ component
corresponds to $(x-x_c)^{i\beta}$ and the left-travelling wave, while the
$I_{-i\beta}$ component corresponds to the right-travelling wave. These
two must be balanced appropriately in order to form a localised standing
wave. Lastly, recall
 that $k_y>0$ and thus $x_c<0$.  The sign of $x_c$ is important in what
 follows as it determinines
the asymptotic forms of the Bessel functions far from corotation.

Consider first the limit $x\to \infty$. The asymptotic forms of the two
solution components can be derived using
\begin{align*}
I_{i\beta}(X)= \frac{1}{\sqrt{2\pi\,X}}\,\left(e^{X}+ i e^{-\pi\beta-X}
\right),
\end{align*}
where $X=x-ix_c$ (Abramowitz and Stegun 1972). To force the solution $u$ to
decay when $x$ is large and positive,
the coefficient of $\text{exp}(x)$ must be set to zero, and this gives us the
relation
\begin{equation}\label{eig1}
b_1/a_1 =-1.
\end{equation}
This combination means that $u$ is proportional to ${K_\nu(x-ix_c)}$, the
associated Bessel function of second kind (mentioned above).

Consider next the other side of the critical point in the
 limit $x\to-\infty$. The correct asymptotic form to use in this
region is
\begin{align*}
I_{i\beta}(X)=\frac{e^{-\pi\beta}}{\sqrt{2\pi\,X}}\,\left(-e^{-\pi\beta+X}+
ie^{-X}\right),
\end{align*}
because the argument of $X$ approaches $-\pi$ from below (Abramowitz and
Stegun 1972). In order to satisfy
the boundary condition for large negative $x$, 
the coefficient of $\text{exp}(-x)$ must be
zero. But this condition does not lead to a constraint on $x_c$. 
Instead, we obtain
\begin{equation} \label{eig2}
b_1/a_1 = e^{-2\pi\beta},
\end{equation}
which is incompatible with
Eq.~\eqref{eig1}.  For large $\beta$ the ratio (\ref{eig2}) is tiny. 
This very small ratio compensates for the severe damping
that the left-travelling wave $I_{i\beta}$ suffers in this region (see
Eq.~\eqref{phaser}).

The requirement for two different linear combinations precludes the
existence of a continuous analytic solution. There are no normal
modes: when $\sigma\neq 0$ the solution is not continuous at $x=0$,
when $\sigma=0$ the solution is continuous, but possesses undefined
derivatives at $x=0$, a difficulty that can not be resolved whatever the sign of
$\sigma$.

Let us attempt to understand this mathematical difficulty in more physical
terms.  The reason for the two incompatible linear combinations is the
heavy damping caused by the singularity near $x=0$: this strong
damping prevents the simultaneous satisfaction of the two boundary
conditions.  The correct combination of waves in the region $x>0$
violates the boundary condition in the  $x<0$ region and vice versa.
For example, suppose we take the correct combination of components in
$x>0$ and analytically continue this solution into the $x<0$ region
(in so doing passing
above the singularity, as always). When our solution passes through
the vicinity of corotation, the (right-going) $I_{-i\beta}$ component
will abruptly pick up an `amplification' factor  $e^{\pi\beta}$ (because
we are proceeding `backward in time' with respect to this wave), while
the $I_{\nu}$ component picks up a damping factor $e^{-\pi\beta}$ (as we
are proceeding forward in time). As a consequence, the amplitudes of the
two waves will differ by $e^{-2\pi\beta}$. But this large difference will
upset the balance necessary to satisfy the decaying boundary condition at
negative infinity, which requires the two components to be of comparable
amplitude (at the very least). Thus as $x$ becomes large and negative,
$u$ will blow up because it will be dominated by the coefficient of $e^{-x}$
that issues from
$I_{\nu}$. If we take the correct combination of waves in $x<0$ and
follow them into $x>0$ the same problem is encountered.

For small $\beta$ (less than $1/2$), the order of the Bessel functions
$\nu$ is real and each solution component picks up a phase shift,
rather than an amplitude change, when it is followed through corotation
(see Eq.~\eqref{phaser}). But this phase shift is also sufficient to
prohibit the formation of normal modes as before: if we remove the
divergent component of the solution when $x\to\infty$, the phase jump
across $x=0$ means we are unable to remove the divergent component of
the solution at $x\to-\infty$.

Though there exist no analytic eigenfunctions, when $\sigma=0$
a piecewise solution can be devised which is continuous at $x=0$ but whose
derivative, as already noted, is
undefined at this point: 
\begin{equation}\label{singm}
u_\text{s} = \Theta(x)\,u_+(x) + \Theta(-x)\,u_-(x),
\end{equation}
where $\Theta(x)$ is the Heaviside step function and the right and left
functions are
\begin{align}\label{up}
u_+ =& a_1^* x^{1/2}\left[I_{\nu}(x) - I_{-\nu}(x)   \right], \\
u_-=&  b_1^* x^{1/2} \left[I_{\nu}(x) + e^{2\pi i \nu} I_{-\nu}(x)  \right],\label{um}
\end{align}
where $a_1^*$ and $b_1^*$ are two undetermined constants.
 In fact, because of translational symmetry,
 we can define such a neutral singular `mode' at each $x$. As an ensemble this
 set of structures (understood as `weak' or `distributional' solutions) defines a
continuous spectrum, because each radius is a singularity for at least one mode.
 Associated with the non-normal operators which
 appear regularly in shear flow problems, 
such `solutions' aid in 
the evolution of initial data and thus only have
physical meaning as kernels in the integral expressions of $u$ (see for
 example Schmid and Henningson, 2001).
 In particular,
the continuous spectrum is often responsible for
 transient growth effects. The initial
 value problem we address in Appendix C, and in Section 5
 where we show the relationship between the continuous spectrum and shearing waves.

In summary, this simple demonstration shows the main problem trapped
inertial waves suffer if they are to form around their corotation
singularity.  The shearing sheet model, though probably inadequate to
fully describe inertial waves in real relativistic disks, here suffices
to illustrate the principal physical and mathematical issues  at play. In
short, the absence of normal modes stems from the selective wave damping
at the corotation radius. As a consequence, potential standing waves,
composed of both right and left-travelling waves can never be calibrated
so that they decay in the far field on either side of corotation. (In
Appendix A these results are generalised to a slightly incompressible
fluid, and in Appendix B the influence of viscosity is considered.)

\section{Absence of trapped inertial waves in a semi-global model}

In this section, we present a WKBJ calculation in a more realistic
semi-global model of a disk.   In the limit of large $k_z$, the main
result of the previous section is recovered.
The analysis is similar to, but somewhat simpler than,
the development in K02. Nevertheless, it is still involved and the
reader may skip directly to Section 5 without loss of continuity.

Our strategy is to first assume that trapped standing waves grow,
$\sigma>0$, then to determine their approximate profiles near the two
Lindblad resonances (subsections 4.2.1 and 4.2.2 and 4.3) and near
the corotation resonance (subsection 4.2.3), and finally to match
these three solutions in the WKBJ regions which border their domains of
validity (subsection 4.4). The matching conditions supply an eigenvalue
equation \eqref{eigveq} for the growth rate $\sigma$ of
the trapped modes. We approximately solve this equation, as in K02,
and determine that in fact $\sigma<0$. We conclude, as a consequence,
that no such trapped modes can exist.

\subsection{The wave-mode equation}

We begin with an incompressible disk in cylindrical geometry $(r,\phi,z)$
orbiting with frequency profile $\Omega(r)$. If the isothermal approximation
is used, the dynamics remain essentially unaltered (though more
mathematically involved).  The system admits an equilibrium state
characterised by the (constant) density $\rho_0$ and the flow $\u=
r\Omega\,\ephi$.  A small perturbation taking the form $\propto
\text{exp}(im\phi+ik_z z - i\omega t)$ is governed by the following
linearised equations,
\begin{align}
& i \widetilde{\omega} u_r' + 2 \Omega u'_\phi = \d_r h', \\
& i \widetilde{\omega} u_\phi' - \kappa^2/(2\Omega)\,u_r' = i (m/r) h', \\
& i \widetilde{\omega} u_z' = i k_z h', \\
& 0= u_r'/r + \d_r u_r' + i (m/r) u_\phi' + i k_z u_z', 
\end{align}
where $h=p'/\rho_0$ and $\widetilde{\omega}=\omega-m\Omega$. 
The epicyclic frequency is defined through $\kappa^2(r)=
2\Omega(r\d_r\Omega+ 2\Omega)$ but, in order to approximately account for
general relativistic effects near a black hole, $\kappa(r)$ may take other forms
(see Okazaki et
al.~1987, Kato 1990).

This set can be manipulated into a simple second order ODE for $u_r'$, which
may be significantly simplified (as in K02) by
assuming that the mode oscillates rapidly in comparison
with the radial variation of the background equilibrium. If 
 $$\d_r \ln \kappa,\,\d_r\ln\Omega \sim K, $$
 where $K$ is some characteristic
 wavenumber, then $K\ll k_z$. Also we let
$$ u_r'/r \ll  \d_r u_r' \sim k_z\,u_r', \qquad k_z \gg m/r.$$
If we normalize the unit of length so that $K=1$,
the wave-shape equation is approximated by
\begin{equation} \label{wmeq}
\d_r^2 u_r' + \epsilon^2 f(r,\omega)\, u_r' = 0
\end{equation}
where 
\begin{equation} \label{fff}
 f(r,\omega) = \frac{\kappa^2 -
  \widetilde{\omega}^2}{\widetilde{\omega}^2},
\end{equation}
and $\epsilon=k_z/K$ is a large dimensionless parameter.
Equation \eqref{wmeq} is analogous to Eq.\eqref{Bessel1} in Section 2 and
  Eq.~(20) in K02.
 For notational ease, hereafter we drop the prime and $r$ subscript on $u_r'$.

Equation \eqref{wmeq} possesses
 two turning points $R_1$ and $R_2$, the inner and
outer Lindblad resonances, which are defined through $$ \wt=-\kappa,
\qquad \wt = \kappa $$ respectively. These two radii circumscribe a
resonant cavity in which low frequency standing waves may localise. In
addition, there exists the corotation singularity $R_c$ defined through
$\wt=0$. As noted by Kato (2001b, 2002), for nonzero $m$, this point
usually falls between  $R_1$ and $R_2$, though there exists a small
interval of $\omega$ for which this is not the case 
(see Perez et al.~1997). 
 This special circumstance is not investigated here and we
assume $R_{1r}<R_{cr}<R_{2r}$ (where the $r$ subscript indicates real
part).  Because the resonant cavity contains the corotation singularity,
neutral modes are prohibited. Consequently, if trapped normal modes are
to exist, they must occur in growing/decaying pairs.

We set $\omega=\omega_r+ i\sigma$ where $\omega_r$ and $\sigma$ are real
and $0<\sigma\ll \Omega$.  The analysis is hence limited to slow growing
modes. As a consequence, $R_1$, $R_2$ and $R_c$ gain small imaginary
parts. These can be approximated, to leading order in $\sigma$, by
\begin{align}
& R_{1i}= \frac{\sigma}{m\Omega'_1-\kappa'_1}, \\
& R_{2i}= \frac{\sigma}{m\Omega'_2+\kappa'_2},\\ 
& R_{ci}= \frac{\sigma}{m\Omega'_c},
\end{align}
where a prime indicates differentiation with respect to $r$, and a
subscript of $1$, $2$, or $c$ indicates evaluation at $r= R_{1r}$,
$R_{2r}$, or $R_{cr}$ respectively. Even if $\kappa$ is assumed to possess a
turning point near $R_{cr}$ (as it will in the inner part of a general
relativistic disk) it is easy to see that $R_{1i}$, $R_{2i}$, and $R_{ci}$
are negative for all $m$ and that $$ R_{1i}< R_{ci}<R_{2i}.$$

\subsubsection{`Rossby-wave term'}

Before continuing we draw attention to the `Rossby-wave term' that will
become important when
some of the scaling assumptions are relaxed --- namely, if
 we let $k_y\sim k_z$, and suppose there exist localised
regions where the orbital frequency varies rapidly. If this were the case, the function
$f$ would pick up the following term,
\begin{equation}
\frac{(m/r)\,\d_r(\kappa^2/\Omega)}{2\,k_z^2\,\widetilde{\omega}}.
\end{equation} 
This simple pole may give rise to `Rossby wave instability' in
two-dimensional ($k_z=0$) dynamics (Lovelace et al.~1999, Li et al.~2000),
though for the large $k_z$ scalings we employ here Rossby modes do not
appear.
In particular, near corotation $\widetilde{\omega}=0$ the Rossby wave term is
 dominated by the double pole in \eqref{fff}. 
Having said that, it is conceivable
 that when $k_z$ is small and the orbital frequency possesses a localised
 region
 of exceptionally strong shear, instability could arise. If so there
 will exist
a critical $k_z$ above which the double pole
 dominates the dynamics and consequently eliminates potential Rossby wave
 instability. We do not attempt calculate this critical vertical 
wavenumber here, though we do investigate similar behaviour
 in Section 6 with respect to
 the Papaloizou-Pringle instability in a slender torus (see also Sternberg et
 al.~2008).
 In any case, realistic
 $\Omega$ and $\kappa$ profiles are unlikely to exhibit such strong variation;
 if Rossby-wave instability is to occur it will probably have to rely on steep
  density
 gradients (which we have excluded with the assumption of
 incompressibility).
\footnote[1]{We thank
   the reviewer for bringing this issue to our attention.}

\subsection{Approximate solutions near the Lindblad and corotation resonances}

\subsubsection{Inner Lindblad radius}
Near $r=R_1$ we can let $f= \alpha_1(r-R_1)$ to leading order, where $\alpha_1=
\d_r f$ evaluated at $r=R_1$.
 The wave-mode equation subsequently takes the form of a complex
 Airy's equation
$$ \d_r^2 u +\epsilon^2\alpha_1 (r-R_1)\,u =0,$$
with solution
\begin{equation}
u= a_1\,\text{Ai}[\beta_1(R_1-r)] + b_1\,\text{Bi}[\beta_1(R_1-r)],
\end{equation}
in which $\text{Ai}[z]$ and $\text{Bi}[z]$ are the two Airy functions, $a_1$ and
$b_1$ are two complex constants, and 
\begin{equation}
\beta_1= \epsilon^{2/3}\,\alpha_1(-\alpha_1)^{-2/3}.
\end{equation}
The constant $\alpha_1$ will possess a small imaginary component whose sign
plays an important role in negotiating the Stokes phenomenon we encounter
later.
 For small $\sigma$ we can approximate the imaginary part of $\alpha_1$
with the compact expression
\begin{equation} \label{Ima1}
\text{Im}(\alpha_1)= -\frac{2\sigma}{\kappa_1}\left(\frac{\d}{\d
r}\ln\left[\frac{\kappa'-m\Omega'}{\kappa}\right]\right)_1\,.
\end{equation}
As earlier, the subscript $1$ indicates evaluation at $r=R_{1r}$.
 If the logarithm is expanded it is easy to see that the right hand side is
 positive. Then we can write
$$\beta_1= \epsilon^{2/3}\alpha_1^{1/3}\,e^{2\pi i/3}.$$
Because the imaginary part of $\alpha_1$ is positive, the complex argument of
$\beta_1$ is a little greater than $2\pi/3$

\subsubsection{Outer Lindblad radius}
Near $r=R_2$, we can approximate the solution as in the previous section.
As before, we obtain a complex Airy's equation 
\beq \d_r^2 u +\epsilon^2\,\alpha_2 (r-R_2)\, u  =0\eeq
where  $\alpha_2 = \d_r f$, evaluated at $r=R_2$. 
The solution takes the form;
 \begin{equation}
u = a_2\,\text{Ai}[\beta_2(R_2-r)] + b_2\,\text{Bi}[\beta_2(R_2-r)],
\end{equation}
in which $a_2$ and
$b_2$ are two complex constants, and 
\begin{equation}
\beta_2= \epsilon^{2/3}\,\alpha_2(-\alpha_2)^{-2/3},
\end{equation}
The imaginary part of $\alpha_2$ to leading order is
\begin{equation} \label{Ima2}
\text{Im}(\alpha_2)= \frac{2\sigma}{\kappa_2}\left(\frac{\d}{\d
r}\ln\left[\frac{\kappa'+m\Omega'}{\kappa}\right]\right)_2\,.
\end{equation}
It can be shown that the right hand side
is negative for all $m$ for realistic $\kappa$ and $\Omega$ profiles. Now,
$$\beta_2= -\epsilon^{2/3} (-\alpha_2)^{1/3}.$$
Because the real part of $\alpha_2$ is negative, the complex argument of
$\beta_2$ is a little less than $-\pi$.

\subsubsection{Corotation radius}
When $r$ is close to $R_c$ we obtain the approximate equation
$$ \d_r^2 u +k_z^2\,\alpha_c (r-R_c)^{-2}\,u =0,$$
with
$$ \alpha_c = \frac{\kappa^2}{m^2(\Omega')^2} $$
evaluated at $r=R_c$.
This admits a solution of the form
\begin{equation}
u = a_c\,(r-R_c)^{1/2+\beta_c} + b_c\,(r-R_c)^{1/2-\beta_c},
\end{equation}
where $a_c$ and $b_c$ are complex constants and, to leading order in large
$\epsilon$,
\begin{equation}
\beta_c = \epsilon\,(-\alpha_c)^{1/2}.
\end{equation}
We denote this solution hereafter by $u_c$.

The imaginary part of $\alpha_c$ to leading order in small $\sigma$ is given by
\begin{equation} 
\text{Im}(\alpha_c) = \frac{2\kappa_c^2
  \sigma}{m^3(\Omega'_c)^3}\,\left(\frac{\d}{\d r}
\ln\left[\frac{\kappa}{\Omega'}\right]\right)_c\,.
\end{equation}
This can be demonstrated to be negative for all $m>0$. Thus
$$\beta_c= i\,\alpha_c^{1/2}\,\epsilon.$$

\subsection{Behaviour far from the resonant cavity}

We require the trapped mode to decay outside its resonant cavity, which is
circumscribed by $R_1$ and $R_2$. By applying these two boundary conditions
the two unknown constants $b_1$ and $b_2$ can be determined.

\subsubsection{The region $r<R_{1r}$ }

First we inspect the inner region of the disk, $r<R_{1r}$, far from the
resonant cavity. In this region we have $|\epsilon^{2/3}(R_1-r)|\gg 1$ for
sufficiently large $\epsilon$, which supplies the following
 asymptotic forms for the Airy functions:
\begin{align*}
&\text{Ai}[\beta_1(R_1-r)] = \left[\beta_1(R_1-r)\right]^{-1/4}\,\left(e^{-\xi}
+ i\,e^{\xi}\right) \\
&\text{Bi}[\beta_1(R_1-r)] = \left[\beta_1(R_1-r)\right]^{-1/4}\,\left(i\,e^{-\xi}
+ e^{\xi}\right), 
\end{align*}
where
\begin{equation}
\xi= \frac{2}{3}\left[\beta_1(R_1-r)  \right]^{3/2}.
\end{equation}
Because of Stokes phenomenon the argument of $\beta_1(R_1-r)$ is key and we
must use \eqref{Ima1} to choose the correct asymptotic expression\footnote[1]{See for
example http://functions.wolfram.com/Bessel-TypeFunctions/AiryAi.}.
 This also
tells us that $\text{Re}(\xi)<0$. So in order to obtain a decaying $u$ we must
eliminate the $\text{exp}(-\xi)$ terms and keep the
$\text{exp}(\xi)$ terms. This can be accomplished if $b_1=i\,a_1$. 
The solution near the inner Lindblad resonance we now denote by $u_1$
and write as
\begin{equation}
u_1= a_1\left\{\,\text{Ai}[\beta_1(R_1-r)] + i\,\text{Bi}[\beta_1(R_1-r)] \, \right\}.
\end{equation}

\subsubsection{The region $r>R_{2r}$}

We turn now to the outer region of the disk far from the resonant cavity and
with $r>R_{2r}$. Here for sufficiently large $\epsilon$ we have $
|\epsilon^{2/3}(R_2-r)| \gg 1$, and the Airy functions are approximated by
\begin{align*}
&\text{Ai}[\beta_2(R_2-r)] =
\frac{1}{2\sqrt{\pi}}\left[\beta_2(R_2-r)\right]^{-1/4}\,e^{-\zeta} \\
&\text{Bi}[\beta_2(R_2-r)] =
\frac{1}{\sqrt{\pi}}\left[\beta_2(R_2-r)\right]^{-1/4}\,\left( \frac{i}{2}\,e^{-\zeta}
+ e^{\zeta}\right),
\end{align*}
with
\begin{equation}
\zeta = \frac{2}{3}\left[\beta_2(R_2-r) \right]^{3/2}.
\end{equation}
The correct asymptotic form follows from Eq.~\eqref{Ima2}, which also tells us
that $\text{Re}(\zeta)>0$. To ensure decaying solutions we zero all the
$\text{exp}(\zeta)$ terms, which corresponds simply to $b_2=0$. The
solution near the outer Lindblad resonance is subsequently denoted by $u_2$
and written as
\begin{equation}
u_2 = a_2\,\text{Ai}[\beta_2(R_2-r)].
\end{equation}

\begin{figure}
\begin{center}
\scalebox{.55}{\includegraphics{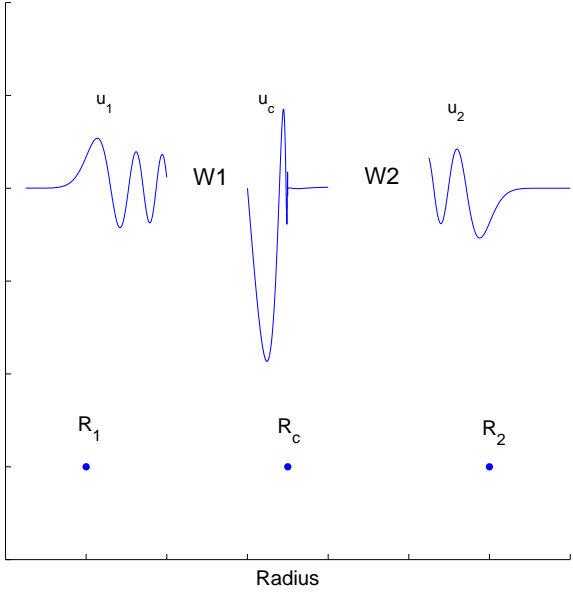}}
\caption{A schematic diagram of the three solutions $u_1$, $u_c$, and $u_2$
  in their regions of validity: near the inner Lindblad, corotation, and outer
  Lindblad radii respectively. These are represented by $R_1$, $R_c$, and
  $R_2$. Between the solutions are the regions $W_1$ and $W_2$ where the solution
  takes its WKBJ form.}
\end{center}
\end{figure}

\subsection{WKBJ matching}

In Fig.~1 the preceding three solutions are shown schematically,
each separated from
its neighbour by either region $W_1$ or $W_2$. In these two regions we assume $u$
takes its WKBJ form:
\begin{align*}
& u = a_W\,f^{-1/4}\,\text{exp}\left( i\,\epsilon\int f^{1/2} dr\right) \\
&\hskip3cm + b_W\,f^{-1/4}\,\text{exp}\left(-i\,\epsilon\int f^{1/2} dr\right).
\end{align*}
Our task now is to extend solutions $u_1$, $u_2$, and $u_c$ into regions $W_1$
and $W_2$ using the above ansatz and then to obtain matching conditions
which set $\sigma$.

\subsubsection{The $u_1$ solution in $W_1$}
Suppose that $R_1$ and $R_c$ are sufficiently spaced (or $\epsilon$
sufficiently large) so that
$|\epsilon^{2/3}(R_1-r)|\gg 1$ in region $W_1$. The correct asymptotic
form for $u_1$ here is 
$$ u_1= \frac{a_1}{\pi^{1/2}}\left[\beta_1(R_1-r)  \right]^{-1/4}\,\left(e^{-\xi} +
  i\,e^{\xi}\right),$$
which can be manipulated into approximate WKBJ form. First we treat $r$ as a
  complex variable, by simply adding to it a
  small imaginary component equal to $iR_{1i}$. Then we do the following:
\begin{align*}
\phi_1 &\equiv \int_{R_1}^r f^{1/2}\,dr,\\
    &= \int_{R_1}^r \left[\alpha_1(r-R_1)\right]^{1/2}\,dr\,\, +\,\,
    \mathcal{O}[(r-R_1)^{5/2}],\\
    &= i\xi\,\epsilon^{-1} \,\,+\,\,\mathcal{O}[(r-R_1)^{5/2}].
\end{align*}
Here the path of integration is simply a straight horizontal
line in the complex plane connecting $R_1$ to $r$.
We now have
\begin{equation} \label{W1}
u_1 \approx \widetilde{a}_1\,f^{-1/4}\,\left( e^{i\epsilon\phi_1} +
i\,e^{-i\epsilon\phi_1} \right)
\end{equation}
in the region $W_1$, where $\widetilde{a}_1$ is a new complex constant.

\subsubsection{The $u_2$ solution in $W_2$}
Suppose that $|\epsilon^{2/3}(R_2-r)|\gg 1$ in region $W_2$. The correct asymptotic
form for $u_2$ here is 
$$ u_2=
\frac{a_2}{2\pi^{1/2}}\left[\beta_2(R_2-r)\right]^{-1/4}\left(e^{-\zeta}+i\,e^{\zeta}\right).$$
As earlier, we treat $r$ as a complex variable but this time add a small
imaginary component equal to $iR_{2i}$. Next we define
\begin{align*}
\phi_2 \equiv \int_{R_2}^r f^{1/2}\,dr,
\end{align*}
where the integration contour is a straight horizontal line joining $R_2$ to $r$. To
leading order it can be demonstrated that $\epsilon\phi_2=-i\zeta$ and so
\begin{equation} \label{W2}
u_2 \approx \widetilde{a}_2\,f^{-1/4}\,\left( e^{-i\epsilon\phi_2} +
i\,e^{i\epsilon\phi_2}\right)
\end{equation}
in region $W_2$.

\subsubsection{The $u_c$ solution in $W_1$ and $W_2$}
So far our analysis has largely followed that of K02, but now we will
deviate a little. The corotation radius introduces a problematic branch point in
the asymptotic definition of $u_c$ and as a consequence it is obliged to take two
different
WKBJ expressions, one in region $W_1$ and another in region $W_2$. This subtlety
appears to be incompletely treated in the two papers by Kato. Our approach
here is relatively clunky but is unambiguous.

The $u_c$ solution is rewritten as
$$ u_c= \sqrt{r-R_c}\left( a_c\, e^{\beta_c\ln(r-R_c)} +
  b_c\,e^{-\beta_c\ln(r-R_c)} \right).$$
Now to account for the solution in $W_1$ we define
\begin{align*}
\phi_{3} = \int_{R_{cr}+iR_{1i}}^r f^{1/2} dr,
\end{align*}
where the contour is the straight line connecting the point
$R_{cr}+ i\,R_{1i}$ to $r$ in the region $W_1$. Here $r$ has taken the small imaginary
component $i\,R_{1i}$. It can be shown to leading order in region $W_1$
\begin{align*}
i\epsilon\phi_{3} = -\beta_c\,\ln(r-R_c) +
\epsilon\alpha_c^{1/2}\left[i\ln(i\Delta_1)-\pi\right],
\end{align*}
where $\Delta_1= R_{ci}-R_{1i}$. Note the large damping/amplification
 factor $- \epsilon\pi\alpha^{1/2}_c$ which appears to have been missed in
 K02. In addition, in $W_1$ we have
$$ (r-R_c)^{1/2} \approx i\alpha_c^{1/4}\,f^{-1/4}.$$
Similarly in $W_2$ we define
\begin{align*}
\phi_{4} = \int_{R_{cr}+iR_{2i}}^r f^{1/2} dr.
\end{align*}
 The contour of integration is the straight line connecting the point
$R_{cr}+ i\,R_{2i}$ to $r$ in the region $W_2$. Here, however, $r$ has taken a
small imaginary component $i\,R_{2i}$. To leading order
\begin{align*}
i\epsilon\phi_{4}= \beta_c\,\ln(r-R_c) - \epsilon\alpha_c^{1/2}i \ln(-i\Delta_2),
\end{align*}
with $\Delta_2= R_{ci}-R_{2i}$. For $r$ in $W_2$, 
$$ (r-R_c)^{1/2} \approx \alpha_c^{1/4}\,f^{-1/4}.$$

In summary, the WKBJ solution in region $W_1$ is
\begin{align}
&u_c\approx i\,f^{-1/4}\,\left\{\,\, \widetilde{a}_c\,
  \text{exp}\left( \epsilon\sqrt{\alpha_c}[i\ln(i\Delta_1)-\pi]
  -i\epsilon\phi_{3} \right) \right. \notag \\
& \hskip1cm \left. 
+\widetilde{b}_c\,
  \text{exp}\left( -\epsilon\sqrt{\alpha_c}[i\ln(i\Delta_1)-\pi
]+i\epsilon\phi_{3}\right)\,\, \right\}. \label{W1c}
\end{align}
The WKBJ solution in region $W_2$ is
\begin{align}
&u_c\approx f^{-1/4}\left\{\,\, \widetilde{a}_c\,
  \text{exp}\left(i\sqrt{\alpha_c}\epsilon\ln(-i\Delta_2)+\epsilon\phi_{4}\right)
\right. \notag \\
& \hskip1cm \left. 
+\widetilde{b}_c\,\text{exp}\left(-i\sqrt{\alpha}\epsilon\ln(-i\Delta_2)-i\epsilon\phi_{4}
\right)\,\,\right\}. \label{W2c}
\end{align}

\subsubsection{Asymptotic matching}
First the two solutions \eqref{W1} and \eqref{W1c} are compared in region $W_1$.
We define 
$$\Psi_1  \equiv \int_{R_1}^{R_{cr}+iR_{1i}} f^{1/2}\,dr = \phi_1-\phi_{3}  $$
and then equate coefficients of $\text{exp}(i\epsilon\phi_1)$ and of
$\text{exp}(-i\epsilon\phi_1)$. This yields the two constants $\widetilde{a}_c$ and
$\widetilde{b}_c$
in terms of $\widetilde{a}_1$, which without loss of generality we let equal
to 1.
We have, as a consequence,
\begin{align} \label{match1}
\widetilde{a}_c &= \text{exp}\left\{ \epsilon\alpha_c^{1/2} [\pi-i\ln(i\Delta_1)] -
i \epsilon\Psi_1
\right\}, \\
\widetilde{b}_c &= -i\,\text{exp}\left\{
-\epsilon\alpha_c^{1/2}[\pi-i\ln(i\Delta_1)] +i\epsilon\Psi_1
\right\}.\label{match2}
\end{align}

Now we compare solutions \eqref{W2} and \eqref{W2c} in region $W_2$. We define
$$\Psi_2  \equiv \int_{R_{cr}+iR_{2i}}^{R_2} f^{1/2}\,dr =
\phi_4-\phi_{2}, $$
and equate coefficients of $\text{exp}(i\epsilon\phi_2)$ and
$\text{exp}(-i\epsilon\phi_2)$. 
With Eqs \eqref{match1} and \eqref{match2} we obtains two equations, one which
gives us
$\widetilde{a}_2$ and finally the eigenvalue equation for $\omega$. Once
$\widetilde{a}_2$ has been eliminated and, after some algebraic manipulation,
the eigenvalue equation takes the form
\begin{equation} \label{eigveq}
\Psi_1-\Psi_2 + \alpha_c^{1/2}\left[\ln\left(\frac{-\Delta_1}{\Delta_2}\right) +i\pi
\right]=\pi\,n
\end{equation}
where $n$ is some integer. In the next subsection we attempt to solve this
equation approximately.

\subsection{Approximate solution to the eigenvalue equation}
For perhaps all realistic $\kappa$ and $\Omega$,
 the integral functions $\Psi_1$ and $\Psi_2$ cannot be computed
 analytically. Instead, following K02 and K03a, we obtain approximations by using
 the asymptotic forms of $f$ near $R_1$, $R_2$, and $R_c$. Doing so certainly
 introduces some level of error, though the essential qualitative points we
 make should remain unaltered.

The $\Psi_1$ integral is decomposed as follows
\begin{align*}
 \Psi_1 &= \int_{R_1}^{R_{cr}+iR_{1i}} f^{1/2}\,dr, \\
        &= \int_{R_1}^{R_3} f^{1/2}\,dr + \int_{R_3}^{R_{cr}+iR_{1i}}
        f^{1/2}\,dr,
\end{align*}
where $R_3$ is an intermediate point on the line connecting $R_1$ and
$R_{ci}+iR_{1i}$. In the first integral we approximate $f$ by its asymptotic
form near $R_1$, i.e.\ $f= \alpha_1(r-R_1)$, while in the second we
use its form near $R_c$, $f= \alpha_c (r-R_c)^{-2}$. The integrals are then
 straightforward to calculate and we find
$$ \Psi_1 \approx \frac{2}{3}\alpha_1^{1/2}(R_{3r}-R_{1r})^{3/2}
+\alpha_c^{1/2}\ln\left(\frac{R_c-R_3}{i\Delta_1}\right).$$
An analogous procedure furnishes us with
$$ \Psi_2 \approx -\frac{2}{3}(-\alpha_2)^{1/2}(R_{2r}-R_{4r})^{3/2} 
+ \alpha_c^{1/2}
\ln \left( \frac{R_4- R_c}{-i\Delta_2}\right),$$
where we have introduced the intermediate point $R_4$ which lies on the line
connecting $R_2$ and $R_{cr}+iR_{2i}$. 

With these expressions, the eigenvalue equation \eqref{eigveq} becomes the
more manageable
\begin{align}
& \frac{2}{3}\alpha_1^{1/2}\Delta R_1^{3/2}
 +\frac{2}{3}(-\alpha_2)^{1/2}\Delta R_2^{3/2}  \notag \\ 
& \hskip2cm +i\alpha_c^{1/2}\pi +  \alpha_c^{1/2}\ln\left(\frac{\Delta R_3 +
 i\Delta_1}{\Delta R_4 -i\Delta_2}\right) =
 \pi\,n, \label{finaleigveq}
\end{align}
where
\begin{align*}
\Delta R_1 &= R_{3r}-R_{1r},
&\Delta R_2 = R_{2r}-R_{4r},\\
\Delta R_3 &= R_{cr}-R_{3r},
&\Delta R_4 = R_{4r}-R_{cr}.
\end{align*}
Next we use the assumption that $\sigma$ is small and expand $\alpha_1$, $\alpha_2$,
$\alpha_c$ in powers of $\sigma$,
\begin{align*}
 \alpha_1&= \alpha_{11}+\alpha_{12}\,i\sigma +\mathcal{O}(\sigma^2), \\
 -\alpha_2&= \alpha_{21}+\alpha_{22}\,i\sigma+\mathcal{O}(\sigma^2), \\
 \alpha_c&=\alpha_{c1}+\alpha_{c2}\,i\sigma+\mathcal{O}(\sigma^2),
\end{align*}
where the $\alpha$ coefficients on the right sides are all real and positive
except for $\alpha_{c2}$, (see Section 4.2). Note that K02 neglects these
higher order corrections in the analogous terms of his analysis; however,
these corrections will be important in the imaginary component of
Eq.~\eqref{finaleigveq}. 
We next take the imaginary part of
Eq.~\eqref{finaleigveq} but neglect the last term on the left side. This term
contributes terms proportional to 
$\sigma\ln(\Delta R_1/\Delta R_2),$
$\sigma/\Delta R_1$, and $\sigma/\Delta R_2$
all of which will be subdominant to the other terms 
if $R_1$, $R_2$, and $R_c$ are sufficiently well separated so that the $\Delta
R_i$ are large and of the same order.
Finally, we can solve for $\sigma$,
\begin{equation}
\sigma= \frac{-3\pi \alpha_{c1}^{1/2}}{\alpha_{12}\alpha_{11}^{-1/2}\Delta
  R_1^{3/2} +\alpha_{22}\alpha_{21}^{-1/2}\Delta R_2^{3/2}}.
\end{equation}
Thus $\sigma$ is negative. 

However, a negative $\sigma$ contradicts the initial assumption that
$\sigma>0$ (so
crucial in negotiating the Stokes phenomenon). If, on the other,
hand we begin the analysis with the assumption $\sigma<0$ we finish with the
opposite result: $\sigma$ must be positive. The only resolution to the impasse
is to take $\sigma=0$. But this means that the modes will possess undefined
derivatives. The conclusion is
that there exist no trapped 3D normal modes for such WKBJ waves.  

Before we continue, it should be noted that the negative results of Sections 3
and 4 and Appendix A pertain to specific cases associated with certain
approximations: e.g.\  the local shearing sheet model and incompressible and slightly
incompressible fluids, and when in a more realistic geometry, the WKBJ limit
of large $k_z$. They are hence not proofs for the non-existence of \emph{all}
trapped 3D inertial waves near corotation. Together these results, however,
offer a good argument that this is indeed the case.

\section{Shearing waves and the continuous spectrum}

In this section we briefly demonstrate the dynamical behaviour
 of the continuous spectrum introduced  in the shearing sheet analysis of
 Section 3. This is accomplished by inverting the temporal Fourier transform,
\begin{equation}\label{inv1}
 U(x,t) = \int_{\Gamma} u(x,\omega)\,e^{-i\omega t}\,d\omega,
\end{equation}
where $u(x,\omega)$ is the solution to the modal problem and $\Gamma$ is the
appropriate integration contour in the complex $\omega$ plane. It is assumed that
$\omega$ is real except near any singularity, at which point the integration
path $\Gamma$ will deviate above the singularity.

 For a
typical initial value problem, in which the initial condition is stipulated,
 we would construct $u$ using a Greens function. Usually such problems throw
 up complicated integrals that are difficult to solve, and the analysis is usually
limited
 to the large time asymptotic regime (Booker and Bretherton 1967,
 Watts et al.~2004, for example). In Appendix C we undertake such an analysis
 and show that localised initial conditions evolve towards decaying shearing inertial
 waves (for a description of shearing waves see
 Johnson and Gammie 2005, Balbus and Hawley 2006).
This long time behaviour issues from the collective influence of
  the corotation singularities. To emphasise this point, we present a
 slightly different calculation below in which each member of the continuous
 spectrum is summed equally. Doing so allows the inversion integral to be
 solved analytically and thus compared with the exact shearing wave solution of
 Johnson and Gammie (2005).

\subsection{Shearing waves}
Let us start with the integral \eqref{inv1} and immediately change the
 integration variable to
\begin{equation}\label{coord1b}
 \theta= k_\perp\,x + \frac{k_\perp\,\omega}{q\Omega k_y}.
\end{equation}
This is identical to the $x^*$ introduced earlier, but in the following it is denoted
differently in order to avoid confusion with $x$. The new coordinate measures the distance
between $x$ and the corotation point specified by $\omega$. In addition, a
dimensionless time coordinate is introduced
\begin{equation}\label{coord2}
\tau=  \left(\frac{q\Omega k_y}{k_\perp}\right)\,t
\end{equation}
 We hence obtain
\begin{equation} \label{inversion}
U= \left(\frac{q\Omega k_y}{k_\perp}\right)\,e^{i\,k_x(t) x}\,
\int_{\widetilde{\Gamma}} \hat{u}(x,\theta)\,e^{-i\tau \theta} \,d\theta,
\end{equation}
where $\widetilde{\Gamma}$ is a contour in $\theta$ space,
$u(x,\omega)=\hat{u}(x,\theta)$, and 
 the time-dependent (shearing) radial wavenumber is
$$k_x(t)= q\Omega k_y\,t.$$
A simple shift in time introduces a constant $k_x'$ into the above expression,
and will render $k_x(t)$ in the familiar form used in Balbus (2003).
 The integral \eqref{inv1} now represents not a sum over all the $\omega$ 
but, for a given $x$ location, a sum over all the corotation points
that influence the dynamics.

Before we specify the
form of $u$ the basic structure of the solution should be appreciated. First, we have a
shearing wave contribution from the $\text{exp}[ik_x(t)\,x]$. Second,
 this spatial dependence
will be modulated by a possible $x$-dependent factor issuing from the
integral in \eqref{inversion}, the details of which are linked to the initial
condition. 
This integral will also set the time-dependence of the solutions' amplitude. For
large times the amplitude decays like $t^{-3/2\pm \nu}$ (just
as in Booker and Bretheton 1967). The dominant contributions to the integral
in this case come from the combined criticial radii (the corotation points).

To better clarify the action of the continuous spectrum, we treat an equal sum
of its component singular modes; there is hence no modulation arising from
localised initial conditions.  Consequently, we set $\hat{u}(x,\theta)=u_s(\theta) $,
where $u_s$
is defined in Eq.~\eqref{singm}. Now the spatial structure of the solution
 becomes a pure shearing wave with the integral in
Eq.~\eqref{inversion}
 depending
on time only through $\tau$. We let this integral equal the amplitude function $A(\tau)$,
which we now calculate. The amplitude function involves a linear combination
of the two integrals:
\begin{align*}
& A_1= \int_0^\infty
\sqrt{\theta}\left[I_\nu(\theta)-I_{-\nu}(\theta)\right]\,e^{-i\tau
\theta}\,d\theta,\\
& A_2= \int_{-\infty}^0 \sqrt{\theta}\left[I_\nu(\theta)+e^{2\pi
    i\nu}I_{-\nu}(\theta)\right]\,e^{-i\tau \theta}\,d\theta.\\
\end{align*}
Both integrals are proportional to
$$ \int_0^\infty \sqrt{\theta}\, K_\nu (\theta)\,e^{-i\tau \theta}\,d\theta,$$
 which can be evaluated in closed form in terms of hypergeometric
functions (Gradshteyn and Rhyzik 1963). Finally, after some tedious algebraic
manipulation using a chain of functional identities, one
can express the time-dependent amplitude of the shearing wave as
\begin{align*}
A(\tau)&=  c_1\,
F(\tfrac{3}{4}-\tfrac{1}{2}\nu,\tfrac{3}{4}+\tfrac{1}{2}\nu;\tfrac{1}{2};-\tau^2)\notag
\\
&\hskip2cm  + d_1\tau
  F(\tfrac{5}{4}-\tfrac{1}{2}\nu,\tfrac{5}{4}+\tfrac{1}{2}\nu;\tfrac{3}{2};-\tau^2),
\end{align*}
where $c_1$ and $d_1$ are two constants and $F(a,b,c;z)$ is
a Gauss hypergeometric function, often represented as ${}_2F_1$. 
This amplitude is exactly that of a pure shearing wave computed
directly from the governing equation \eqref{firststep} rendered in
 shearing coordinates:
\begin{equation}
\d_\tau^2 A + \beta^2(1+\tau^2)^{-1}\,A=0,
\end{equation}
(for the derivation see
 Johnson and
Gammie 2005 or Balbus and Hawley 2006). 

 The main point to take away is that
 inertial shearing waves are equivalent to the continuous spectrum issuing
 from
the corotation singularities. This simple relationship has not been generally
appreciated in the astrophysics community,
 but is quite natural (Craik and Criminale, 1986). A shearing wave understood this way is but a sum
of structures $u_s$ each tightly localised to its critical radius and
convected perfectly upon the background shear, like tracers.

\section{Instability in a slender torus}

The previous sections have investigated the role of the corotation point in
 an extended disk. The analysis was local and the disk-boundaries 
were sent to infinity. Consequently, trapped modes could not arise.
In this section a different scenario is attacked: an inviscid slender torus, the radial
 boundaries of which play the essential part. Specifically, we
examine the effect the corotation point has upon the incompressible three-dimensional
 Papaloizou-Pringle instability (Papaloizou and Pringle 1984, 1985,
Drury 1985, Goldreich et al.~1986). 

The classical model of this
global hydrodynamical instability is two-dimensional, in which case the corotation
point is not a singularity of the governing equations. The mechanism of
instability proceeds in this case from the efficient communication of angular
 momentum from the inner boundary to the outer via wave motions across the
 corotation point. If, however, the mode in question posesses vertical
structure, the corotation point, being singular, will impede this transfer 
and either inhibit growth or prohibit the formation of normal modes entirely (as in
the
previous sections). In the following we characterise this
 phenomenon for general $q$. Note that our results complement those obtained
 by Sternberg et al.~(2008) who examine a similar problem with fixed boundary walls.

\subsection{Mathematical set-up}

We consider the shearing sheet approximation to a narrow incompressible fluid
annulus with no equilibrium
vertical structure. The centre of the sheet is anchored at the centre of the
annulus $r_0$ so that the fluid occupies the radial region circumscribed by
$|x|\leq s$, where $2s$ is the total radial width of the annulus.
 The Eulerian perturbation equations are Eqs~\eqref{lin1}-\eqref{lin4},
 and we draw our general solution for $u_x'$ for a given
 $(k_y,k_z,\omega)$ mode from
Eq.~\eqref{Bessel1}. 

The appropriate boundary conditions for a narrow annulus is that the
Lagrangian enthalpy perturbation must vanish at the free surfaces $x=\pm
s$. This condition may be represented by
\begin{equation} 
h'+ \Delta x\, \frac{d h_0}{d x} = 0,
\end{equation}
where 
$$ \Delta x = \frac{i u_x'}{q\Omega k_y x + i\sigma} $$
is the radial displacement at the boundary. The Eulerian enthalpy perturbation
is supplied by 
$$ h'= \frac{i\widetilde{\omega}}{k_\perp^2}\,\d_x u_x' +
i(2-q)\Omega\frac{k_y}{k_\perp^2}\,u_x',$$
and the radial gradient of the equilibrium enthalpy $h_0$ is 
$$ \frac{d h_0}{d x}=-(2q-3)\Omega^2 \,x.$$
This enthalpy gradient aids gravity negate the centrifugal force when the
rotation law is non-Keplerian.

In this section we consider modes that are corotating with the centre of the
annulus, so that $\omega_r=0$, but which possess nonzero $\sigma$. We also
scale space by $k_\perp^{-1}$ as in Section 2.3, while introducing the
scaled growth rate
$$ \overline{\sigma}=
\frac{k_\perp}{q\,k_y}\left(\frac{\sigma}{\Omega}\right),$$
(which in Section 2.3 is equal to $-x_c$). The boundary condition can now be
expressed neatly as
\begin{align} 
&q^2\,(x+i\overline{\sigma})^2\,\frac{d u_x'}{d x} +\{ q(2-q)(x+i\overline{\sigma})
  \notag \\
 &\hskip2.5cm   -(2q-3)(1+k_z^2/k_y^2)x\}u_x'
  =0, \label{boundary}
\end{align}
which must hold at $x=\pm s$. The expression for $u_x'$ is
\begin{equation*}
u_x'= A\sqrt{x +i\overline{\sigma}}\,I_{\nu}(x+i\overline{\sigma}) +
B\sqrt{x+i\overline{\sigma}}\,I_{-\nu}(x+i\overline{\sigma}),
\end{equation*}
where $A$ and $B$ are constants to be determined, and $\nu$ can be computed
from \eqref{nu}. Once $A$ and $B$ are eliminated from the two equations we
arrive upon the dispersion relation for $\overline{\sigma}$, which is a function
 of the three dimensionless parameters, $q$, $k_z/k_y$,
and $s$.

\subsection{Growth rates}

 Unfortunately, the dispersion relation is transcendental in
$\overline{\sigma}$. In the interest of readability, the full expression
 is banished to
Appendix D.
In general, the growth rates must be computed numerically.
Nevertheless, some analytic progress can be made if it is assumed that the
width of the annulus is small relative to the typical lengthscale of a $(k_y,
k_z)$ mode,
i.e.\ $|s|\ll 1$. If the annulus was much wider, especially if the mode's
Lindblad resonances fitted into the domain, communication between the two boundaries
via incompressible motions would fail and the instability would evaporate.
In addition, we let the growth rates scale
 as $\overline{\sigma}\sim s$ (Balbus, 2003). This permits the scaling
$ I_\nu(i\overline{\sigma} \pm s)\sim s^\nu, $
 which helps us fish out the most significant terms in the dispersion relation. 
To leading order it can then be boiled down to
\begin{align} \label{reddisp}
&2q^2(2-q)\,f(\sigma_1) \sigma_1^2 + 2q^2\,i\,(2q-3)\,\nu\,
g(\sigma_1)\,\sigma_1\notag \\
& \hskip1.7cm + 
f(\sigma_1)\left[3(3-q^2) + \frac{k_z^2}{k_y^2}\,(3-2q)^2\right] = 0,
\end{align}
where we have set $\overline{\sigma}= s\sigma_1+\mathcal{O}(s^2)$ and where
\begin{align*}
f(\sigma_1)&= (i\sigma_1-1)^{2\nu}-(i\sigma_1+1)^{2\nu}.\\
g(\sigma_1)&=  (i\sigma_1-1)^{2\nu}+(i\sigma_1+1)^{2\nu}.
\end{align*}
Though the reduced dispersion relation cannot be solved
analytically for $\sigma_1$, it yields two significant results.

 First, we can establish the two-dimensional
stability criterion easily. By setting $k_z=0$, and consequently $\nu=1/2$, we
get the twinned growing and decaying modes $$ q\sigma_1 = \pm \sqrt{3(q^2-3)} $$
in agreement with Papaloizou and Pringle (1985) (see also Balbus 2003).
 The instability criterion is thence $q>\sqrt{3}$.

\begin{figure}
\begin{center}
\scalebox{.55}{\includegraphics{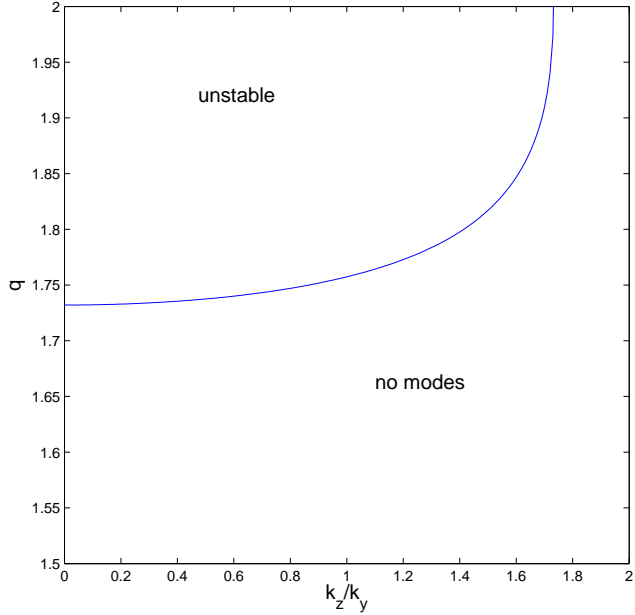}}
\caption{The marginal stability curve in the parameter space $(k_z/k_y,q)$
   given by \eqref{marginal}. The
  region above it is unstable. The region bounded by it and the vertical
  $k_z/k_y=0$ axis circumscribe parameters for which no modes exist.}
\end{center}
\end{figure}

Second, it yields the curve of marginal stability for general
 three-dimensional modes once we
  set $\sigma_1=0$. The curve is
described by
\begin{equation}\label{marginal}
q_\text{crit} = \frac{3}{4(k_z/k_y)^2-3}\,\left( 2(k_z/k_y)^2-\sqrt{3-(k_z/k_y)^2}
\right).
\end{equation}
 The curve of critical
$q$ as a function of $k_z/k_y$ is plotted in Fig.~2. There we can
see clearly that the greater the vertical pitch of the wavevector, the greater the
critical $q$ necessary
 for instability. 

As expected, the most unstable modes are two-dimensional. Three-dimensional
modes, on the other hand, must contend with the singularity at corotation $x=0$, which
 interferes with the mechanism of instability.
 As discussed in Section 2.4, waves with small wavevector pitch,
 $k_z/k_y\leq
q\Omega/(2\kappa)$ (i.e.\ $\beta\leq 1/2$), 
 pick up a phase shift upon crossing this point.
 But the singularity is far more destructive for modes
with greater pitch, $k_z/k_y>q\Omega/(2\kappa)$ (i.e.\ $\beta>1/2$) which are severely damped.
 No normal modes (growing or otherwise) of the latter class
are possible in the slender torus.
 This is also true for some low $k_z/k_y$ modes which only suffer the phase
shift. Generally, however, such $\beta\leq 1/2 $ modes do exist and occur as
growing/decaying pairs. They are situated in the region above the marginal
curve in Fig.~2. 
Their growth rates are lower than the equivalent 2D mode
at the same $q$ because the phase shift renders the transfer of angular momentum across
corotation less inefficient.  This is illustrated particularly well in
the eigenfunction profiles in the next subsection.

\subsection{Eigenfunctions}

To examine the effect of the corotation singularity on the structure of the
unstable modes we fix $q=1.8$ and $s=0.1$ and subsequently vary the ratio
$k_z/k_y$ from $0$ to the value at which modes cease to exist (near $1.5$).
 The eigenproblem Eq.~\eqref{disp} is solved numerically by a
Newtown-Raphson method.

\begin{figure}
\begin{center}
\scalebox{.55}{\includegraphics{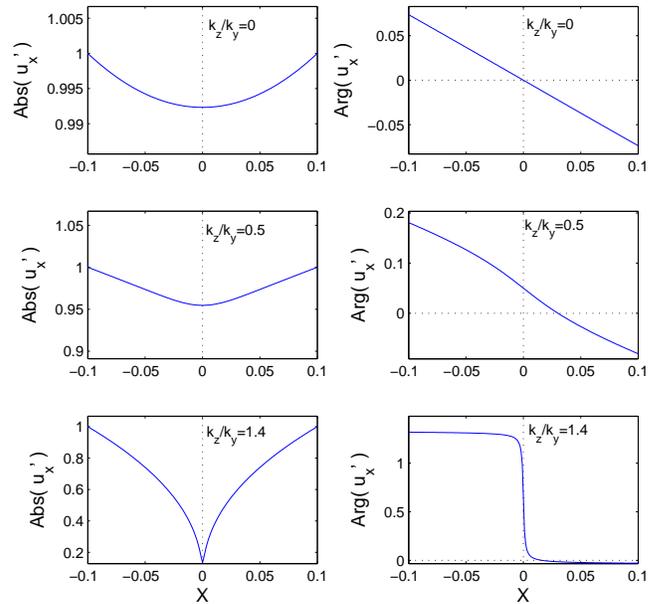}}
\caption{The modulus and argument of the
 $u_x'$ eigenfunction's radial structure for $q=1.8$, $s=0.1$, and various $k_z/k_y$.
Panels (a) and (b) show the (two-dimensional) eigenfunction when
$k_z/k_y=0$. The growth rate here is  $\sigma=7.60\times10^{-2}\,\Omega$.
 Panels (c) and (d) show the case when $k_z/k_y=0.5$. 
 Note that the presence of the corotation
  singularity breaks the radial symmetry. As a consequence,
the unscaled growth rate is less: we compute $\sigma=
6.23\times10^{-2}\,\Omega$.
Finally, panels (e) and (f) show an eigenfunction with $k_z/k_y=1.4$,
 a value which
  positions the mode very near the stability boundary sketched in Fig.~3. The unscaled
  growth rate is $\sigma= 9.91\times 10^{-4}\,\Omega$. Note the near
  discontinuity in the argument at corotation. Motions on either side of this
  point are out of phase by a quantity near $\pi/2$, and thus the transfer of
  angular momentum and energy from the inner boundary to the outer is severely
  impeded.
 The eigenfunctions are normalised so that the
 maximum of $|u_x'|$ is 1.}
\end{center}
\end{figure}

In Fig.~3, the top two panels (a) and (b) exhibit the Eulerian radial velocity
component ($u_x'$) 
of the classical
two-dimensional mode ($k_z=0$) as a function of dimensionless $x$. 
Both the modulus and argument are presented. The most
important features
here are the relatively small variation in radius ($x$), which is of order
$s^2$, and particularly the small change in phase from the inner edge to the
outer edge.

In the next two panels, Figs 3c and 3d, a three-dimensional mode is plotted with $k_z/k_y=0.5$. The
mode grows slower than the previous case, and the symmetry of its structure is
broken. In Figs 3e and 3f the $k_z/k_y=1.4$ case is presented, which is near
criticality. Any value of $k_z/k_y$ much larger does not return a modal solution.
The salient feature here is the near discontinuity in 
$\text{Arg}(u_x')$ at the corotation radius. Motions on either side of this
point
 are out of phase by a quantity
that approaches $\pi/2$. This represents the
limit of severe
damping of energy and angular momentum transfer and thus zero growth.

\section{Conclusion}

Axisymmetric inertial waves can become trapped in the inner regions of disks
orbitting black holes, forming standing modes (Okazaki et al.~1987). In certain
cases a small set of
non-axisymmetric inertial waves can do the same (Perez et al.~1997). However,
most 3D non-axisymmetric inertial waves possess a corotation singularity
within their resonant cavity and,
as a consequence, these waves will have difficulty forming trapped standing motions.

This idea was demonstrated with a simple local model, the shearing sheet, which
crystallises the important physical and mathematical points
thus allowing a straightfoward interpretation of the problem. Its conclusions are
bolstered by a WKBJ analysis in a more realistic semi-global model in
cylindrical geometry in Section 4, and a compressible analysis in Appendix A.
 The central point is that a trapped
standing wave must carefully balance its two component travelling
waves in order to satisfy the two decaying boundary conditions far away from
the corotation region; at the very least, these two travelling waves must possess
comparable amplitudes. But the corotation singularity acts as a powerful wave
absorber, heavily damping one of the waves relative to the other and hence
destroying the necessary balance. If one boundary condition is satisfied the
other must be violated.

 In summary, non-axisymmetric 3D inertial waves should play little direct role
 in QPO models based on diskoseismology, though they may be implicated in parametric
instabilities, where they can act as `intermediaries' transferring energy between two
trapped
axisymmetric waves (for example, Ferreira and Ogilvie 2008).
More generally, the viability of diskoseismology rests on the interactions between
 trapped
 inertial waves and their turbulent environment. Recent local and global simulations, in particular, have shown
 that trapped modes struggle to emerge 
from MRI-induced turbulence (Arras et al.~2006, Reynolds and Miller 2009).

The corotation singularity, while forbidding discrete normal modes, on the
  other hand generates a continuous set
  of neutral singular `modes', a continuous spectrum.
 We show that an ensemble of these singular structures
  corresponds exactly to the inertial shearing waves computed by Johnson and
  Gammie (2005) and Balbus and Hawley (2006), and in Appendix C we use them to
  solve the initial value problem for localised initial conditions in the
  asymptotic limit of large time. This provides an alternative, and
  analytically tractable, interpretation which helps deepen our understanding
  of shearing waves, which appear naturally in shearing box simulations and
  may be important in planet-disk interactions (Balbus 2003, Balbus and Hawley
  2006, Shen et al.~2006).

Lastly, unstable 3D modes in a simple model of a slender torus were studied in
the context of corotational damping. As expected, 3D modes of small pitch
$k_z/k_y$ grow at slower rate than their 2D counterparts on account of the
impeded angular momentum transfer at the corotation radius, which is a
singularity when $k_z\neq 0$. Modes of greater
pitch do not exist at all. This simple example analysis fleshes out rather
nicely the ideas put forward
in Li et al.~(2003). Moreover, it shows that astrophysical tori which are
unstable to the incompressible Papaloizou and Pringle instability will be dominated
by 
the two-dimensional unstable modes above all others.

\section*{Acknowledgements}
The authors would like to thank the anonymous reviewer for helpful comments
which much improved the paper. H.~N.~L thanks Christopher Heaton for helpful
and encouraging advice
on the continuous spectrum, and also to Gordon Ogilvie who pointed out
important flaws
in our treatment of viscosity in an earlier draft.
This work was has been supported
 by a grant from the Conseil R\'egional de l'Ile de France.

\appendix

\section{Departures from incompressiblity}

This appendix reinforces the result of Section 3 by investigating the role of
compressibility. The most salient effect of compressibility in our problem 
is the `leakage' of inertial waves through the walls of their
confining potential wells. We establish whether this
tunneling
 can circumvent the strong constraints imposed by the boundary conditions and
 consequently
 permit trapped normal mode solutions.
 Recently Ferreira and Ogilvie (2008) numerically
demonstrated that a trapped axisymmetric inertial mode can tunnel through its
confining
barrier and emerge on its other side
 as a small-amplitude p-mode. The effect
seems small but may be important in
the normal mode formation (or, rather, non-formation).

 In light of this we undertake a generalisation of
the analysis of Section 3
 to a `slightly incompressible' fluid;
that is, we define a small parameter associated with the sound speed and
expand to its first order. The incompressible analysis of the main paper is
subsequently interpreted as the zeroth order approximation. As before, we
attempt to describe trapped normal r-modes but again we find that no such
solutions exist.

\subsection{Governing equations}

We return to the set of linearised equations for a small disturbance
but now use the full continuity equation 
$$\d_t \rho +
\u\cdot\nabla\rho=-\rho\nabla\cdot \u,$$ 
instead of the incompressibility restriction $\nabla\cdot\u=0$ and we assume an
ideal fluid, so that $p= c_s^2 \rho$, where $c_s$ is the (constant) sound
speed. In addition, the tidal potential $\Phi$ picks up a component
representing vertical gravity.

The
vertical equilibrium of the basic state exhibits a Gaussian vertical
structure. Consequently, we assume
a mode structure proportional to
$$ \text{exp}(ik_y y- i\omega
t)\,\text{He}_n[z/(\sqrt{2}H)],$$
where $\text{He}_n$ is a Hermite polynomial of order $n$ and $H=c_s/\Omega$ is the
disk scale height.
In the shearing sheet model this ansatz yields the linearised equations,
\begin{align} 
i\widetilde{\omega}h' &= c_s^2[\d_x u_x'+ i k_y u_y' + i (k_z^2/\widetilde{\omega})\, h'], \\ 
i\widetilde{\omega}u_x' &= -2\Omega u_y' +\d_x h', \\
i\widetilde{\omega}u_y'&=  \tfrac{1}{2}(\kappa^2/\Omega)u_x'+ i k_y h',
\end{align}
in which we have defined $k_z\equiv \sqrt{n}/H$, and eliminated $u_z'$.
 This set of equations can be combined to form an
equation for $u_x'$,
\begin{align*}
&\d_x^2 u_x' + \frac{2q\Omega\,\widetilde{\omega}\,k_y}{ c^2 k_\perp^2
-\widetilde{\omega}^2}\,\d_x
u_x' \\ 
&  +\left\{
  k_\perp^2\left(\frac{\kappa^2}{\widetilde{\omega}^2}-1\right)\left(1-\frac{\widetilde{\omega}^2}{c^2
  k_\perp^2}\right) - \frac{\kappa^2 k_y^2}{\widetilde{\omega}^2} + \frac{q\kappa^2
k_y}{c^2 k_\perp^2- \widetilde{\omega}^2}\right\}u_x'=0.
\end{align*}
This equation exhibits both the regular singular point at corotation,
$\widetilde{\omega}=0$, and the apparent singularities at the vertical
resonances, $ \widetilde{\omega}^2= c_s^2 k_\perp^2$. The latter points, with the
Lindbland resonances $\widetilde{\omega}^2=\kappa^2$, circumscribe the two potential
barriers. We set $\omega=\omega_r+i\sigma$ and change coordinates as in Section
2.  Doing so introduces the new dimensionless
parameter
$$ \delta= (H\,k_\perp)^{-1},$$
which measures the importance of compressible effects. To make
further progress the key
step is to assume this quantity is small and to
expand the mode equation in its powers. The result up to order $\delta^2$ is
\begin{align}
&\d_x^2 u + 2\delta^2\alpha^2\,X\d_x u \notag \\ 
&\hskip1.75cm + \left\{\frac{\beta^2}{X^2}-(1-\gamma) +\alpha^2 X^2 \right\}u
 =0,\label{badarse}
\end{align}
 where $X=x-ix_c$, with $\beta$ and $x_c$ given in Section 2, and
\begin{align*}
\alpha=  q (k_y/k_\perp)\,\delta, \qquad \gamma= \delta^2\kappa^2[q(k_y/k_\perp)^2-1].
\end{align*}
Note that we have dropped the prime and subscript $x$ on $u_x'$.
In addition, we could expand $u$ in small $\delta$ as well, but we retain all
its subdominant components.

The equation \eqref{badarse} is governed by the dimensionless parameters $q$,
$k_y/k_z$,
and $\delta$. By setting $\delta=0$ we recover Bessel's equation
\eqref{Bessel2}. The additional terms alter the nature
of the potential, which is described by the negative of the
term in curly brackets. To leading order, the two forbidden zones are contained, on
the one hand,
between $x=\beta$ and $x= \alpha^{-1}$, and, on the other,
between $x=-\beta$ and $x=-\alpha^{-1}$. In the incompressible
limit we have $\delta\to 0$ and find the barriers extend to infinity, as in
Sections 2 and 3. There are thus three zones in which
 wave-motion can take place. Modes localised primarily to the two regions
 $|x|>\alpha^{-1}$, far away from corotation, we
 identify as p-modes (density waves), and modes localised near corotation
 $|x|<\beta$, we identify as r-modes (inertial waves).

\subsection{Mathematical analysis}
Equation \eqref{badarse} can be solved analytically by  assuming the functional form
$$ u= X^a e^{b X^2} f(X),$$
where $a$ and $b$ are constants we are free to choose. After substitution of
this ansatz into \eqref{badarse} $a$ and $b$ are set to values which zero the
 $x^{-2}$ and $x^2$ terms in the coefficent of $f$. The result is a version
 of Kummer's equation which can be made more explicit by the coordinate
 transform $Y= i\alpha\sqrt{1-\alpha^2}\,x^2$. Then we have
$$ Y\, \d_Y^2 f + (1\pm\nu-Y)\, \d_Y f - \eta f =0 $$
where $\nu$ was introduced earlier in \eqref{nu} and $\eta$ is given through
$$ \eta=
\frac{1\pm\nu}{2}-\frac{i(1+ \alpha^2-\gamma
  )}{4\alpha\sqrt{1-\alpha^2}}.$$
Note the ambiguity in the sign of $\nu$ which arises from a freedom in the
choice of $a$ and which corresponds physically to the direction of the p-mode
solution. Without loss of generality we take the positive
sign (the left-going
p-mode), but remain conscious that results pertaining to it also hold for the
negative case (right-going mode).

 The two independent solutions to Kummer's equations are
 $M(\eta,\nu+1,Y)$ and $U(\eta,\nu+1,Y)$, the confluent hypergeometric
 functions of first and second kind respectively (Abramowitz and Stegun,
1972). The two linearly independent solutions
to the original wave-shape equation \eqref{badarse}, with the correct values of
$a$, $b$, and $Y$ substituted, are hence
\begin{align}
&u_1= X^{1/2+\nu}\,\text{exp}\left(-\tfrac{1}{2}\alpha^2 X^2
  -\tfrac{1}{2}\,i\,\alpha\sqrt{1-\alpha^2}X^2\right)\notag \\
&\hskip2.5cm\times U(\eta,\,1+\nu;\,i\alpha\sqrt{1-\alpha^2}\,X^2\,)\\
&u_2= X^{1/2+\nu}\,\text{exp}\left(-\tfrac{1}{2}\alpha^2 X^2
  -\tfrac{1}{2}\,i\,\alpha\sqrt{1-\alpha^2}X^2\right)\notag \\
&\hskip2.5cm\times M(\eta,\,1+\nu;\,i\alpha\sqrt{1-\alpha^2}\,X^2\,).
\end{align}
\begin{figure}
\begin{center}
\scalebox{.55}{\includegraphics{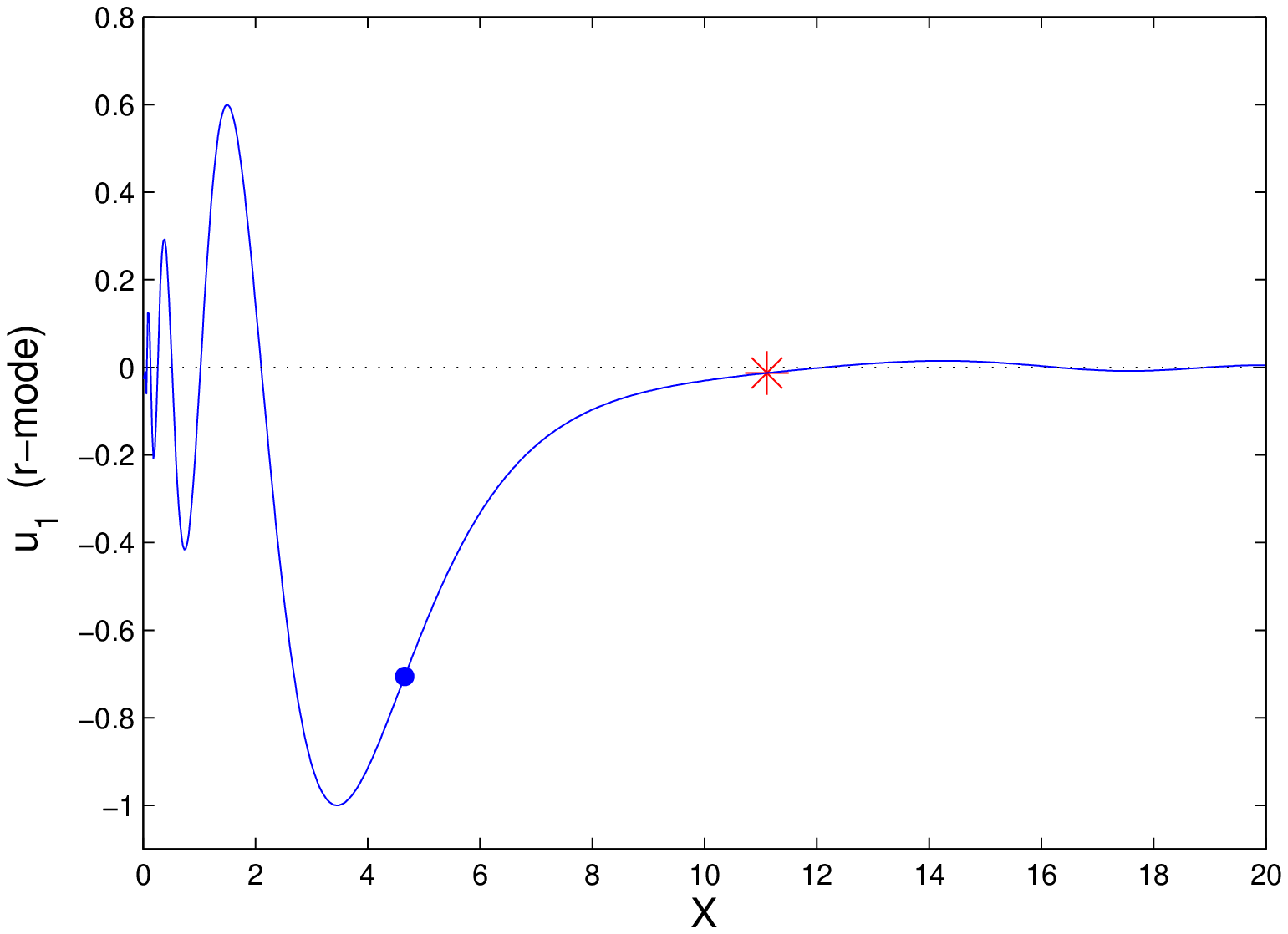}}
\caption{Here we present the real part of $u_1$ for $x>0$ with
  parameters $q=3/2$, $k_z/k_y=7$, and $\delta=0.425$. In addition, we indicate the
location of the
  resonances: the corotation point is at $x=0$; the
  solid circle refers to the Lindblad resonance, $x=4.67 $, and the asterisk refers to the
vertical
  resonance, $x=11.1 $. The forbidden zone is hence delimited by an asterisk and a
  circle. We only plot $u_1$ to identify it as possessing the features of an
  inertial wave.
As is made
  clear in the text, inertial waves do not manifest as normal modes in this problem.}
\end{center}
\end{figure}
\begin{figure}
\begin{center}
\scalebox{.55}{\includegraphics{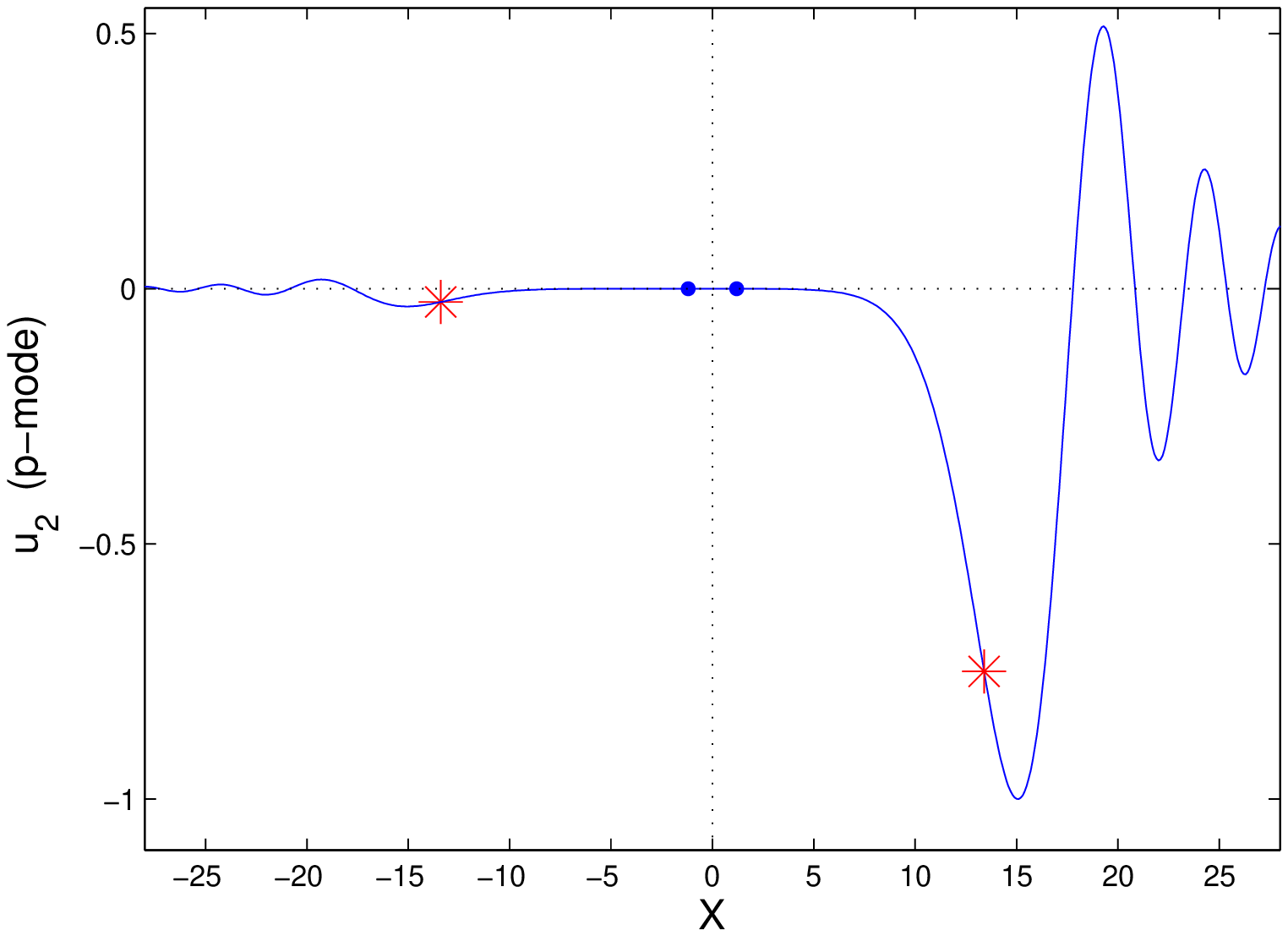}}
\caption{In this figure we plot the real part of the left-travelling p-mode as
  described by $u_2$ as a function of $x$. The parameters are $q=3/2$, $k_z/k_y=
1.75 $, and
  $\delta=0.1$. The Lindblad resonances are at $x=\pm 1.17$ and are denoted by
  solid circles, the vertical resonances
  at $x=\pm 13.4$ and denoted by asterisks, and the corotation point is at $x=0$.}
\end{center}
\end{figure}
We identify solution $u_1$ with the r-mode and $u_2$ with the (left-going) p-mode by
inspection of their profiles (see Figs A1 and A2): $u_1$ is confined almost fully to
the cavity
near corotation with some leakage through the potential barriers; $u_2$ is
confined to the areas outside the potential barriers with some leakage into
the corotation region (which is too small to see). Both solutions
decay speedily for large $|x|$.

\subsection{Absence of trapped inertial waves}

First, it is assumed that $\sigma>0$. Next, consider
the argument of the hypergeometric function in $u_1$, 
$$Y= i\alpha\sqrt{1-\alpha}X^2=
\alpha\sqrt{1-\alpha^2}\left[-2\sigma x + i(x^2-\sigma^2)
\right].$$
Suppose that we following the solution along $x$ towards the corotation point from
the right
(in region $x>0$). When we reach $x=\sigma$ the imaginary part of
$Y$ changes sign but the real part of $Y$ remains negative, which means that
 in the complex $Y$-plane we cross the negative real axis. But the negative
 real axis happens to be
a branch cut for the function $U(\eta,\nu+1,Y)$ (Abramowitz and Stegun 1972). This
means that
$$\lim_{\delta\to 0^+} U(\eta,\nu+1, -Z+ i\delta)\neq \lim_{\delta\to 0^+}
U(\eta,\nu+1, -Z- i\delta) $$
for any $Z$ positive and real. So, in effect, when $\sigma\neq 0$ 
the solution $u_1$ exhibits a
discontinuity when $x=\sigma$ and is not analytic as a result.
  (In fact, $u_1$ proceeds to a different
Riemann surface.) If we set $\sigma=0$ then the discontinuity is removed but
then the solution possesses undefined derivatives at $x=0$ and is thus not analytic
either. Physically, the $u_1$ is a combination of left and right-travelling
p-modes, organised so that $|u_1|\to 0$ as $|x|\to \infty$. But just as in
Section 3, the satisfaction of these two boundary conditions comes at the
price of continuity --- of the function itself or its derivatives.
In summary, there can be no trapped inertial modes even when compressibility
is taken into account.

\subsection{Density waves}

On the other hand, the confluent hypergeometric $M$ of the p-mode $u_2$ is an
entire function and thus
suffers no branch cuts. If there is a small nonzero $\sigma$ the p-mode is
analytic throughout the domain: from Fig.~A2 it impacts on the
potential barrier from the right, decays exponentially in the forbidden region, and
trickles
 into the narrow corotation region before emerging as a smaller amplitude
 p-mode oscillation on the far side of the left-most barrier.
  We suspect the numerical integration undertaken in the
 Appendix A of Li et al.~(2003) produces a profile not dissimilar to $u_2$,
 though it may not capture the function's far-field decay.

\section{Influence of viscosity}

Typically the continuous spectrum gives rise to algebraic
growth or decay. However, even a small amount of dissipation
 will remove the singularity at corotation, and in 
 bounded flows the inviscid continuous spectrum will congeal into a discrete
 but dense set of eigenvalues each associated with a regular eigenfunction
 exhibiting an
 exponential time dependence. 
In the unbounded
shearing sheet, however, the analogous spectrum will remain continuous and its
eigenfunctions singular.

 Though the singularity at corotation is removed
by viscosity, an incident travelling wave should still suffer strong absorption when it
strikes the corotation region. This means the arguments of Section 3, which we
used to dismiss the possibility
that left and right-travelling waves can be combined to form trapped standing
waves, still apply qualitatively. On the other hand, one can
demonstrate the impossibility of constructing a localised time-exponential solution from a sum
of viscous shearing waves, and thus of viscous standing eigenmodes generally (Gordon
Ogilvie, private communication). 
It follows that the unbounded shearing sheet must possess a viscous continuous
spectrum which describes the dynamics of viscous shearing waves.

Just as in the Orr-Somerfeld problem, the effect of a small viscosity in a
disk is most
pronounced at the critical radius -- at corotation.
 A viscous boundary layer develops around this point to mediate the
 absorption of wave energy. From the viscous equations of motion we can derive the
 following sixth-order equation in the dimensionless variables of Section 2.3,
\begin{equation} \label{viscouseq}
\Xi^2\,\d_x^2 u - \Xi^2\,u + \beta^2\,u = 0
\end{equation}
where the operator $\Xi$ is defined by
\begin{equation}
\Xi= (x-ix_c) -i\,\mathcal{R}^{-1}\,(\d_x^2-1),
\end{equation}
which introduces the (modified) Reynolds number
\begin{equation}
\mathcal{R}= \left(\frac{q\Omega}{\nu\,k_\perp^2}\right)\,\left(\frac{k_y}{k_\perp}\right),
\end{equation}
with $\nu$ the kinematic viscosity. From this equation it can be shown that
the internal viscous layer at corotation possesses a width of order
$\mathcal{R}^{-1/3}$ when the Reynolds number is large.
 Approximate solutions very close to
corotation (inside the viscous layer) can be easily derived for WKBJ waves when
 $\beta\sim\mathcal{R}\gg 1$. All six solution decay (or grow) exponentially
through the layer at a (spatial) rate $(\mathcal{R}\,\beta)^{1/3}$. This means the
total reduction in amplitude of an incident WKBJ wave through the entire layer
will be of order $\beta^{1/3}$, which should be contrasted with the rate
$\beta$ that emerges from the inviscid analysis of Section 2.4. Thus,
viscosity allows incident waves to penetrate corotation with greater success,
though the damping will still be sufficient to preclude standing eigenmodes.

\section{The initial value
  problem}

This appendix offers a more 
formal treatment of the initial value problem than in Section 5.
We return to the original linearised equations \eqref{lin1}-\eqref{lin4} and
take Fourier transforms in space and time. The temporal Fourier transform is
defined by
$$ \hat{f} = \int_0^\infty f\,e^{i\omega t}\,dt. $$
It follows that
\begin{align*}
&-i\widetilde{\omega} \hat{u}_x - 2\Omega \hat{u}_y = -\d_x \hat{h} + u_x^0, \\
&-i\widetilde{\omega} \hat{u}_y + \frac{\kappa^2}{2\Omega}\, \hat{u}_x = -i k_y
\hat{h} +
u_y^0, \\
&-i\widetilde{\omega} \hat{u}_z = -ik_z \hat{h} + u_z^0
\end{align*}
with the incompressibilty condition. 
It is assumed that the initial
conditions, $\u^0$ and $h^0$,
 are analytic and obey the far field decaying boundary conditions.
Like in Section 3 this set is
reduced to a
single equation for $\hat{u}_x$. With the space dimension scaled by
$k_\perp^{-1}$ and with the transformed velocity field scaled by $\Omega$,
 we obtain
\begin{equation}\label{inhomo}
\d_x^2 \hat{u}_x + \left(\frac{\beta^2}{(x-x_c)^2}-1\right)\hat{u}_x=
\frac{1}{(x-x_c)}\,\Lambda(x, x_c),
\end{equation}
where $x_c= -(\omega k_\perp)/(q\Omega k_y)$. Note that this definition is different to
that appearing in Sections 2 and 3. 
Equation \eqref{inhomo} is the inhomogeneous Bessel's equation with
\begin{align*}
\Lambda(x, x_c) &= \frac{i(k_\perp)}{q k_y}\,(\d_x^2 u^0_x - u^0_x) +
\frac{2i}{q^2(x-x_c)} u^0_z.
\end{align*}
To simplify the following mathematics, without altering our conclusions,
$u_z^0$ is set to $0$. Thus $\Lambda=\Lambda(x)$.

\subsection{Greens function solution}

Equation \eqref{inhomo} can be solved using a Greens function, $G(x,y)$, so that
$$ \hat{u}_x= \int_{-\infty}^\infty \frac{\Lambda(y)\,G(x,y)}{y-x_c} dy,$$
in which the integration contour deviates below the singularity at $y=x_c$.
The Greens function must satisfy the equation
$$ \d_x^2 G + \left(\frac{\beta}{(x-x_c)^2}-1\right)G = \delta(x-y),$$
where $\delta(x)$ is the Dirac delta function, while at the same time satisfying the
far field
decaying boundary conditions. A suitable choice is
\begin{align*}
& G(x,y)= \frac{1}{W}\left\{ \Theta(y-x)\cdot u_+(y)\, u_-(x) \right.\\ 
& \hskip4cm \left.+ \Theta(x-y)\cdot  u_+(x)\, u_-(y) \right\},
\end{align*}
where $\Theta(x)$ is the Heaviside step function, $u_+$ and $u_-$ are
\begin{align*}
&u_+= (x-x_c)^{1/2} K_\nu(x-x_c), \\
&u_- =(x-x_c)^{1/2}( I_\nu(x-x_c)- e^{2\pi i\nu} I_{-\nu}(x-x_c) )
\end{align*} 
 and $W$ is
their Wronskian:
$$ W\equiv \left(\frac{d u_+}{d x}\right) u_- \,-\,\left(\frac{d u_-}{d x}\right)u_+
  =(1-e^{2\pi i\nu}).$$ 
We can now write down the solution to Eq.~\eqref{inhomo},
\begin{align}\label{uhat}
\hat{u}_x= A\,u_- + B\,u_+,
\end{align}
where 
\begin{align}\label{Aa}
&A(x,x_c)= \frac{1}{W} \int_x^\infty \,\frac{\Lambda(y) \,u_+(y,x_c)}{y-x_c} \,dy, \\
&B(x,x_c)= \frac{1}{W} \int_{-\infty}^x\,\frac{\Lambda(y)\,u_-(y,x_c)}{y-x_c} \, dy.\label{Bb}
\end{align}

\subsection{Evolution at large times}

Expression \eqref{uhat} is put in the inversion integral to achieve the
full time-dependent solution of $u_x'$.
 From \eqref{inversion} in Section 5 this can be
written as
\begin{equation} \label{inversion2}
U= \left(\frac{q\Omega k_y}{k_\perp}\right)\,e^{i\,k_x(t) x}\,
\int_{\widetilde{\Gamma}} \hat{u}(x,\theta)\,e^{-i\tau \theta} \,d\theta,
\end{equation}
where $\theta=x-x_c$ and $\widetilde{\Gamma}$ is a suitable contour in $\theta$
space. Needless to say, general initial conditions do not yield closed forms
for $A$ and $B$, let alone $U$. However, some progress can be made if we take the
asymptotic limit of large time $\tau\gg 1$. According to the Riemann-Lebesgue
lemma the dominant contribution to the $\theta$ integral occurs when $\theta$
is small because of the $\text{exp}(-i\tau\theta)$ factor. Consequently, we
expand $A$, $B$, $u_+$, and $u_-$ in
$\theta$. For $A$ and $B$ this means $x_c\approx x$, and so $A=A(x)$ and
$B=B(x)$. If $\xi=y-x$ is a
 dummy variable, we get
\begin{align}\label{AA}
A(x) \approx \frac{1}{W}\,\int_0^\infty
\Lambda(x+\xi)\,\xi^{-1/2}\left[I_\nu(\xi)-I_{-\nu}(\xi) \right]d\xi.
\end{align}
and a similar expression for $B$. We assume for the moment that these
leading order expressions for $A$ and $B$ are nonzero and well-defined. 

With $A$ and $B$ no longer depending on
$x_c$ (and hence $\theta$), we can tackle the $\theta$
integral just as in Booker and Bretherton (1967). For small $\theta$ this
integral is proportional to
\begin{align*}
\int_{\widetilde{\Gamma}_e} \left( C(x)\theta^{1/2+\nu}+ D(x)\theta^{1/2-\nu}
\right) e^{-i\tau\theta} d\theta,
\end{align*}
where $\widetilde{\Gamma}_e$ is that portion of the integration contour near
$\theta=0$ and $C(x)$ and $D(x)$ are functions involving combinations of $A$ and $B$.
 An order 1 variable is introduced, $\zeta= i\tau\theta$ which
establishes the basic time dependence of the integral. It can as a consequence
be re-expressed as
$$ \hat{C}(x)\tau^{-3/2+\nu} +\hat{D}(x)\tau^{-3/2-\nu},$$
where the new functions $\hat{C}$ and $\hat{D}$ involve integrals of
the form
$$\int_0^\infty \zeta^{1/2\pm \nu} e^{-\zeta}\,d\zeta,$$
which may be integrated numerically.

Finally, we write down the long time fate of a localised initial
condition. Given $k_y$ and $k_z$  we have
\begin{align}\label{finalic}
u_x'\propto e^{i k_x(t) x+ i k_y y + i k_z
z}\,t^{-3/2}\left(\hat{C}(x)\,t^{\nu}+\hat{D}(x)\,t^{-\nu}\right).
\end{align}
The mode decays algebraically, with oscillatory behaviour when $\nu$ is
imaginary (i.e.\ $\beta>1/2$). The spatial structure of the solution is a
shearing wave localised within an `envelope' defined by
 the functions $\hat{C}$ and $\hat{D}$, which in turn depend on the initial
 condition selected.

\subsection{The behaviour of the $A$ and $B$ functions}

To derive the main result we assumed that the leading order
 terms of $A$ and $B$ are nonzero and well behaved when $\theta$ is small.
In this subsection we attempt to justify these assumptions. In \eqref{AA} we
 expand $\Lambda(x+\xi)$ in a Taylor series around $x$ and truncate at some order
$N$. This
 supplies a reasonable approximation as the 
dominant contribution to the integral comes from near $\xi=0$; moreover
$u_+$ decays rapidly with $\xi$ (like
 $e^{-\xi}$). The first term in the expansion is computed, and we find
\begin{align*}
 A(x) \approx \frac{\sin(\pi\nu)\,\Gamma\left(\tfrac{1-2\nu}{4}\right)
\Gamma\left(\tfrac{1+2\nu}{4}\right)}{\pi\sqrt{2}\,W}\,\Lambda(x)\,+\,\dots
\end{align*}
A similar expression exists for $B$.
 Thus the leading order terms exist and are regular. We can thus
 be assured that generally $A$ and $B$
are both nonzero and well behaved.

\section{Dispersion relation for instability in a three-dimensional
    slender torus}

The full dimensionless dispersion relation which issues from the solvability of
\eqref{boundary}
at $x= \pm s$ can be expressed as
\begin{align}\label{disp}
&\chi_4(\overline{\sigma})\cdot\overline{\sigma}^4
+\chi_3(\overline{\sigma})\cdot\overline{\sigma}^3 
 +\chi_2(\overline{\sigma})\cdot\overline{\sigma}^2\notag\\ &\hskip3cm
+\chi_1(\overline{\sigma})\cdot\overline{\sigma}+\chi_0(\overline{\sigma}) =0,
\end{align}
where
\begin{align*}
&\chi_4=4q^4\left(S_{1+\nu}^{1-\nu}-S_{1-\nu}^{1+\nu}\right), \\
&\chi_3=2iq^3\{\,(q-4-2\nu q)(S_{\nu}^{1-\nu}-S_{1-\nu}^{\nu}) \\
&\hskip3cm +(q-4+2\nu q)(S^{-\nu}_{1+\nu}-S^{1+\nu}_{-\nu})\,\}   , \\
&\chi_2= 8q^2(2-q)(1+\widetilde{k})(S_{-\nu}^{\nu}-S_{\nu}^{-\nu}) \\
&\qquad
+2sq^2(6-8q+q^2+\widetilde{k}(6-4q)-2\nu q^2)(S_{\nu}^{1-\nu}+S_{1-\nu}^{\nu}) \\
&\qquad
-2sq^2(6-8q+q^2+\widetilde{k}(6-4q)+2\nu q^2)(S_{-\nu}^{1+\nu}+S_{1+\nu}^{-\nu}) \\
&\hskip2.5cm +8q^4s^2(S^{1-\nu}_{1+\nu}-S^{1+\nu}_{1-\nu}), \\
&\chi_1=-2iq^2 s\,\{ \,4(2q-3)(1+\widetilde{k})\nu
(S_{-\nu}^{\nu}+S_{\nu}^{-\nu})\\
& \hskip0.75cm+s(12-4q-q^2-4\widetilde{k}(2q-3)+2\nu
q^2)(S^{1-\nu}_{\nu}-S^{\nu}_{1-\nu})
\\
&\hskip0.75cm +s(12-4q-q^2-4\widetilde{k}(2q-3)-2\nu
q^2)(S_{-\nu}^{1+\nu}-S_{1+\nu}^{-\nu})\,\},\\
&\chi_0= s^2\{\, 4[9-3q^2 + \widetilde{k}(3-2q)^2](1+\widetilde{k})
(S_{-\nu}^{\nu}-S_{\nu}^{-\nu})\\
& \qquad+
2q^2s(-6+q^2+\widetilde{k}(4q-6)-2\nu q^2)(S_{1-\nu}^{\nu}+S_{\nu}^{1-\nu})\\
& \qquad -
2q^2s(-6+q^2+\widetilde{k}(4q-6)+2\nu q^2)(S_{1+\nu}^{-\nu}+S_{-\nu}^{1+\nu})\\
& \hskip3cm +4q^4 s^2 (S^{1-\nu}_{1+\nu}-S^{1+\nu}_{1-\nu})\,\},
\end{align*}
in which $\widetilde{k}=(k_z/k_y)^2$ and where we have used the shorthand
\begin{equation*}
S_\nu^\mu= I_\mu(i\overline{\sigma}+s)\,I_{\nu}(i\overline{\sigma}-s).
\end{equation*}
When $\sigma\sim s\ll 1$, Equation \eqref{disp} reduces to \eqref{reddisp} after using the
scaling $$ S_\nu^\mu\sim s^{\nu+\mu}.$$
Consequently, we find only the last three terms in \eqref{disp} contribute to the
leading order balance.


\begin{thebibliography}{40}

\bibitem{AbromStegun72}
Abramowitz, M., Stegun, I.~A., 1972.
\emph{Handbook of Mathematical Functions},
 Dover Press, New York.

\bibitem{AbromoKluz01}
Abramowicz, M.~A., Klu\'{z}niak, W., 2001.
A\&A, 371, L19.

\bibitem{ArrBlaesTurner06}
Arras, P., Blaes, O., Turner, N.~J., 2006.
ApJ, 645, L65.

\bibitem{Balbus2003}
Balbus, S.~A., 2003. ARA\&A, 41, 555.

\bibitem{BalbHawl06}
Balbus, S.~A., Hawley, J.~F., 2006. ApJ 652, 1020.

\bibitem{BookerBretherton67}
Booker, J.~R., Bretherton, F.~P., 1967.
JFM, 27, 513.


\bibitem{Case1960}
Case, K.~M., 1960.
PhFl, 3, 143.

\bibitem{Craik}
Craik, A.~D.~D, Criminale, W.~O., 1986.
RSPSA, 406, 13.

\bibitem{Dr02}
Drazin, P.~G., 2002. \emph{Introduction to Hydrodynamical Stability},
Cambridge Univ.\ Press, Cambridge.

\bibitem{Drury85}
Drury, L.~O'C, 1985. MNRAS, 217, 821.

\bibitem{Barbara08}
Ferreira, B.~T., Ogilvie, G.~I., 2008.
MNRAS, 386, 2297.


\bibitem{GoldLynd65}
Goldreich, P., Lynden-Bell, D., 1965.
MNRAS, 130, 125.

\bibitem{GGN86}
Goldreich, P., Goodman, J., Narayan, R., 1986.
MNRAS, 221, 339.

\bibitem{GR}
Gradshteyn, I.~S., Ryzhik, I.~M., 1963.
\emph{Table of Integrals, Series, and Products}, Academic Press, 
London. 

\bibitem{JohnsonGammie05}
Johnson, B.~M., Gammie, C.~F., 2005.
ApJ, 626, 978.

\bibitem{Kato90}
Kato, S., 1990.
PASJ, 42, 99.

\bibitem{Kato01a}
Kato, S., 2001a.
PASJ, 53, 1.

\bibitem{Kato01b}
Kato, S., 2001b. 
PASJ, 53, L37

\bibitem{Kato02}
Kato, S., 2002.
PASJ, 54, 39.

\bibitem{Kato03a}
Kato, S., 2003a.
PASJ, 55, 257.

\bibitem{Kato03b}
Kato, S., 2003b.
PASJ, 55, 801.

\bibitem{KoryPring95}
Korycansky, D.~G., Pringle, J.~E., 1995.
MNRAS, 272, 618.


\bibitem{LiH00}
Li, H., Finn, J.~M., Lovelace, R.~V.~E., Colgate, S.~A, 2000.
ApJ, 533, 1023.


\bibitem{Lietal03}
Li, L-X., Goodman, J., Narayan, R., 2003.
ApJ, 593, 980. 

\bibitem{Loveless}
Lovelace, R.~V.~E., Li, H., Colgate, S.~A., Nelson, A.~F., 1999.
ApJ, 513, 805.


\bibitem{McClinRem03}
McClintock, J.~E., Remillard, R.~A., 2003.
In \emph{Compact Stellar X-ray Sources}
(eds. Lewin  W.~H.~G.and van der Klis, M.)
 Cambridge Univ.\ Press, Cambridge.


\bibitem{Ogil98}
Ogilvie, G.~I., 1998.
MNRAS, 297, 291.


\bibitem{Okazaki87}
Okazaki, A.~T., Kato, S., Fukue, J., 1987.
PASJ, 39, 457.

\bibitem{PP84}
Papaloizou, J.~C.~B., Pringle, J.~E., 1984.
MNRAS, 208, 721.

\bibitem{PP85}
Papaloizou, J.~C.~B., Pringle, J.~E., 1985.
MNRAS, 213, 799.

\bibitem{Perez97}
Perez, C.~A., Silbergleit, A.~S., Wagoner, R.~V., Lehr, D.~E., 1997.
ApJ 476 589.

\bibitem{ReyMill09}
Reynolds, C.~S., Miller, M.~C, 2009.
ApJ 692, 869.


\bibitem{SchmitHenningsen01}
Schmid, P.~J., Henningson, D.~S., 2001.
\emph{Stability and Transition in Shear Flows}, Springer, New York.


\bibitem{SSG06}
Shen, Y., Stone, J.~M., Gardiner, T.~A,, 2006.
ApJ, 653, 513.

\bibitem{SUmu}
Sternberg, A., Umurhan, O.~M., Gil, Y., Regev, O., 2008.
A\& A, 486, 341.


\bibitem{VishDiam89}
Vishniac, E.~T., Diamond, P., 1989. 
ApJ, 347, 435.

\bibitem{Wagoner99}
Wagoner, R.~V., 1999.
Phys.~Rep., 311, 259.

\bibitem{WAW04}
Watts, A.~L., Andersson, N., Williams, R.~L., 2004.
MNRAS, 350, 927.



\end{thebibliography}
\end{document}